\documentclass[12pt]{article}
\usepackage{amsmath,amssymb}
\usepackage{subfig}
\usepackage{algorithm}
\usepackage{algorithmic}
\usepackage{multirow}
\usepackage{threeparttable}
\usepackage{natbib}
\usepackage{relsize}
\usepackage[colorlinks=true, linkcolor=blue, citecolor=blue, urlcolor=blue]{hyperref}

\renewcommand{\harvardurl}[1]{\textbf{URL:} \url{#1}}
\usepackage{booktabs}
\usepackage{pgf}
\usepackage{tikz}
\usepackage{times}
\usetikzlibrary{arrows,shapes.arrows,shapes.geometric,
	backgrounds,decorations.pathmorphing,positioning,fit,automata}
\usepackage[latin1]{inputenc}
\usepackage{soul,color}
\usepackage{chngcntr}
\usepackage{amsthm}
\usepackage{verbatim}
\usepackage{enumitem}
\usepackage{changepage}
\usepackage{amssymb}
\usepackage{bm}
\usepackage{adjustbox}
\usepackage{makecell}

\definecolor{mygreen}{RGB}{144,241,47}

\usepackage{caption} 
\usepackage{xfrac}
\usepackage{graphicx}
\newcommand{\ind}{\rotatebox[origin=c]{90}{$\models$}}
\date{}
\usepackage{tikz}
\usetikzlibrary{arrows,shapes.arrows,shapes.geometric,
	shapes.multipart,backgrounds,decorations.pathmorphing,positioning,fit,automata}
\tikzset{
	>=stealth',
	true/.style={
		rectangle,
		draw=black, very thick,
		text width=6.5em,
		minimum height=2em,
		text centered,
		fill=gray, opacity = 0.5},
	punkt/.style={
		rectangle,
		rounded corners,
		draw=black, very thick,
		text width=6.5em,
		minimum height=2em,
		text centered},
	est/.style={
		circle,
		draw=black, very thick,
		text centered},
	shade/.style={
		circle,
		draw=black, very thick, fill=gray!50,
		text centered},
	weight/.style={
		circle,
		draw=black, very thick,
		text width=6.5em,
		minimum height=2em,
		text centered},
	pil/.style={
		->,
		thick,
		shorten <=2pt,
		shorten >=2pt,},
	double/.style={
		<->,
		thick,
		shorten <=2pt,
		shorten >=2pt,},
	dash/.style={
		dashed,
		thick,
		shorten <=2pt,
		shorten >=2pt,},
	dashdouble/.style={
		<->,
		dashed,
		thick,
		shorten <=2pt,
		shorten >=2pt,}
}

\newtheoremstyle{note}
{8pt}
{8pt}
{}
{}
{\bfseries}
{:}
{.5em}
{}

\newtheorem{theorem}{Theorem}

\newtheorem{remark}{Remark}
\usepackage[T1]{fontenc}
\newtheorem{assumption}{Assumption}

\newtheorem{corollary}{Collorary}
\newtheorem{proposition}{Proposition}
\allowdisplaybreaks

\usepackage[left=0.6in,right=0.6in,top=1in,bottom=1in]{geometry}
\usepackage{algorithm}
\counterwithout{table}{section}
\usepackage{tikz}

\usepackage{soulpos}
\soulregister\cite7
\soulregister\citep7
\soulregister\ref7
\soulregister\emph7
\soulregister\eqref7
\soulregister\pageref7
\definecolor{mygreen}{RGB}{144,241,47}

\newcommand{\expit}{\text{expit}}

\newcommand{\clr}{\mathrm{clr}}
\newcommand{\clrM}{\mathrm{clrM}}

\newcommand{\beq}{\begin{equation}}
	\newcommand{\eeq}{\end{equation}}
\newcommand{\be}{\begin{eqnarray}}
	\newcommand{\ee}{\end{eqnarray}}

\definecolor{pengcolor1}{RGB}{255,36,0}

\newcommand{\nind}{\not\!\perp\!\!\!\perp}

{\raise0.5ex\hbox{$#1$}\! \left/ \! \lower0.5ex\hbox{$#2$}\right.}

\def\bSig\mathbf{\Sigma}

\usepackage[figuresright]{rotating}

\begin{document}
	
		\def\spacingset#1{\renewcommand{\baselinestretch}%
		{#1}\small\normalsize} \spacingset{1}

	\title{\bf { Inverse probability weighting-based mediation analysis for microbiome data} }

	\author{Yuexia Zhang \\
		Department of Statistics and Data Science, The University of Texas at San Antonio\\
		Jian Wang\\
		Department of Biostatistics, The University of Texas MD Anderson Cancer Center\\
		Jiayi Shen\\
		BeOne Medicines\\
		Jessica Galloway-Pe{\~n}a\\
		Department of Veterinary Pathobiology, Texas A\&M University\\
		Samuel Shelburne\\
		Department of Infectious Diseases, Infection Control, and Employee Health, \\The University of Texas MD Anderson Cancer Center\\
		Linbo Wang\thanks{linbo.wang@utoronto.ca}\\
		Department of Statistical Sciences, University of Toronto\\
		and\\
		Jianhua Hu\thanks{jh3992@cumc.columbia.edu}\\
		Department of Biostatistics, Columbia University
	}
	\maketitle

	\begin{abstract}
\noindent\textbf{Motivation:} Mediation analysis is an important tool for elucidating causal mechanisms in biomedical research and has recently gained increasing attention in microbiome studies. Motivated by a microbiome study of acute myeloid leukemia (AML) patients, we investigate whether the effect of induction chemotherapy intensity  on infection status is mediated through gut microbial taxa. This problem poses substantial methodological challenges due to the high dimensionality, zero inflation, and dependence of microbial mediators, as well as the presence of an exposure-induced mediator--outcome confounder, antibiotic use.\\
\textbf{Results:} To address these challenges, we propose a novel nonparametric identification formula for the interventional indirect effect, a causal mediation estimand that remains identifiable in the presence of exposure-induced mediator--outcome confounding. Building on this identification result, we develop an inverse probability weighting estimator and construct standard normal bootstrap confidence intervals for inference.  Simulation studies demonstrate that the proposed method achieves accurate effect estimation and maintains appropriate type I error rates and power in finite samples. Application to the AML microbiome study provides evidence of a statistically significant interventional indirect effect of induction chemotherapy intensity on infection risk through alterations in the gut microbiome.\\
\textbf{Availability and implementation:}
The R code for the simulation studies is available at
\url{https://github.com/YX-ZHANG-1/CausalMediationMicrobiome} and archived on Zenodo at
\url{https://doi.org/10.5281/zenodo.21960802}.

	\end{abstract}

\spacingset{1.5} 

\section{Introduction}
\label{intro}
The human microbiome has emerged as an important contributor to health and disease, with alterations in microbial composition linked to a wide range of complex conditions, including cardiovascular disease, inflammatory bowel disease, and cancer \citep{koeth2013intestinal, zackular2016manipulation, halfvarson2017dynamics}. 
Advances in next-generation sequencing technologies, such as 16S rRNA and shotgun metagenomic sequencing, have enabled high-resolution characterization of microbial communities and facilitated systematic investigation of the clinical, genetic, and environmental factors shaping the microbiome and its functional impact on human health \citep{chen2016two, wang2020approach}. Increasing evidence further suggests that the microbiome may mediate  the effects of clinical and dietary factors on health outcomes. For instance, allogeneic hematopoietic stem cell transplantation has been shown to disrupt gut microbial communities, with bacterial domination associated with increased infection risk \citep{taur2012intestinal}. Related mediation effects have also been reported for the association between dietary
intake and immune responses or chronic diseases  \citep{sivan2015commensal,koslovsky2020bayesian}.
A deeper understanding of how the microbiome links risk factors to health outcomes may inform novel strategies for disease prevention and intervention through targeted modulation of microbial communities \citep{faith2013long, le2013richness}.

Motivated by a microbiome study of  acute myeloid leukemia (AML) patients conducted at The University of Texas MD Anderson Cancer Center (MD Anderson), we investigate whether gut microbiome features mediate the causal effect of induction chemotherapy (IC) intensity on  infection risk. Infections in cancer patients frequently originate from commensal bacteria \citep{montassier2013recent}, linking infection risk closely  to alterations in the gut microbiome \citep{zitvogel2015cancer}. Patients with AML receiving intensive IC are particularly  vulnerable to microbiota-derived infections \citep{gardner2008randomized}, which contribute to treatment-related morbidity and mortality  \citep{cannas2012infectious}.  Preliminary analyses of AML patients at MD Anderson further showed that baseline microbiome $\alpha$-diversity and treatment-induced changes in diversity were associated with  infections occurring during and shortly after IC \citep{galloway2016role,galloway2020gut}. Together, these findings suggest that alterations in the gut microbiome may mediate the effect of IC intensity on infection risk in patients with AML.

Mediation analysis provides a principled framework for understanding how and why causal effects arise. Classical mediation methods in the social and health sciences are typically formulated within the linear structural equation modeling framework \citep{baron1986moderator,mackinnon2008introduction}. Related approaches have recently been adopted to study mediation effects of microbiome features in human health and disease \citep{zhang2018distance,zhang2021mediation}. However,  mediation effects defined under linear modeling assumptions are model driven and  do not readily generalize beyond linear models. In particular, they are not well suited for settings with binary outcomes, such as infection status in our study.  In contrast, modern causal mediation analysis relies on  nonparametric definitions and identification results. Foundational work by \cite{robins1992identifiability} and  \cite{pearl2001direct}
established nonparametric notions of direct and indirect effects and conditions for their identification. Building on this framework, several methods have been developed to address the high-dimensional and compositional nature of microbiome mediators. These include sparse compositional mediation models for continuous and binary outcomes \citep{sohn2019compositional, sohn2022compositional}, regularized causal mediation approaches using log-contrast and Dirichlet regression models \citep{wang2020estimating}, and methods tailored to zero-inflated microbial mediators \citep{liu2025cramed}.

Despite these advances, existing microbiome mediation methods cannot be directly applied to our AML study due to the presence of exposure-induced mediator--outcome confounding. In particular, antibiotic use both influences gut microbiome composition and infection risk and is itself affected by IC intensity. Thus, ignoring antibiotic use leaves the microbiome--infection relationship confounded, whereas conventional adjustment is problematic because antibiotic use is itself affected by IC intensity. This structure is common in clinical microbiome studies. To address such settings, \citet{vanderweele2014effect} introduced the interventional indirect effect, which remains identifiable in the presence of exposure-induced mediator--outcome confounders and proposed a weighting-based estimation strategy. However, their approach requires explicit modeling of mediator distributions, which is challenging in our AML microbiome study where mediators are high-dimensional, zero-inflated, and  dependent \citep{martin2020modeling}.

To address this challenge, we derive a novel identification formula for the interventional indirect effect that avoids explicit modeling of the conditional distribution of mediators, thereby eliminating the need to specify a potentially complex high-dimensional model. Instead, the proposed approach only requires  modeling the conditional expectation of the binary infection outcome and the conditional distributions of the binary antibiotic-use variable. 
To accommodate the high dimensionality of  microbial mediators, 
we incorporate sparsity-inducing regularization when modeling the  outcome. In addition,  we apply the centered log-ratio (clr) transformation  to account for the compositional nature of microbiome data \citep{aitchison1982statistical, lin2020analysis}. 
The presence of the interventional indirect effect is assessed using a standard normal bootstrap confidence interval \citep{efron1994introduction}.

The remainder of the paper is organized as follows. Section \ref{sec:study} describes the motivating AML microbiome study. Section \ref{sec:method} presents the proposed mediation framework and estimation procedures. Section \ref{sec:simu} evaluates the performance of our proposed method through simulation studies, and Section \ref{sec:apply} applies it to the AML microbiome data. Section \ref{sec:discuss} concludes with a discussion.

\section{The motivating study}
\label{sec:study}

Our work is motivated by an AML microbiome study conducted at MD Anderson, which aimed to understand the role of the microbiome in treatment-related  infectious outcomes. 
The study enrolled $97$ newly diagnosed adult patients with AML  who received IC  at MD Anderson between September 2013 and August 2015 \citep{galloway2020gut}.  Fecal specimens were collected  at baseline (prior to IC initiation) and approximately every 96 hours during IC until neutrophil recovery, yielding 566 samples in total. Microbial DNA was extracted from stool samples, and the V4 region of the 16S rRNA gene was sequenced using the Illumina MiSeq platform. Sequencing reads were clustered into operational taxonomic units at a 97\% similarity threshold at the genus level, producing a taxon-level abundance table for downstream analysis.

\begin{figure}[ht]
	\begin{center}
	\includegraphics[width=0.8\textwidth]{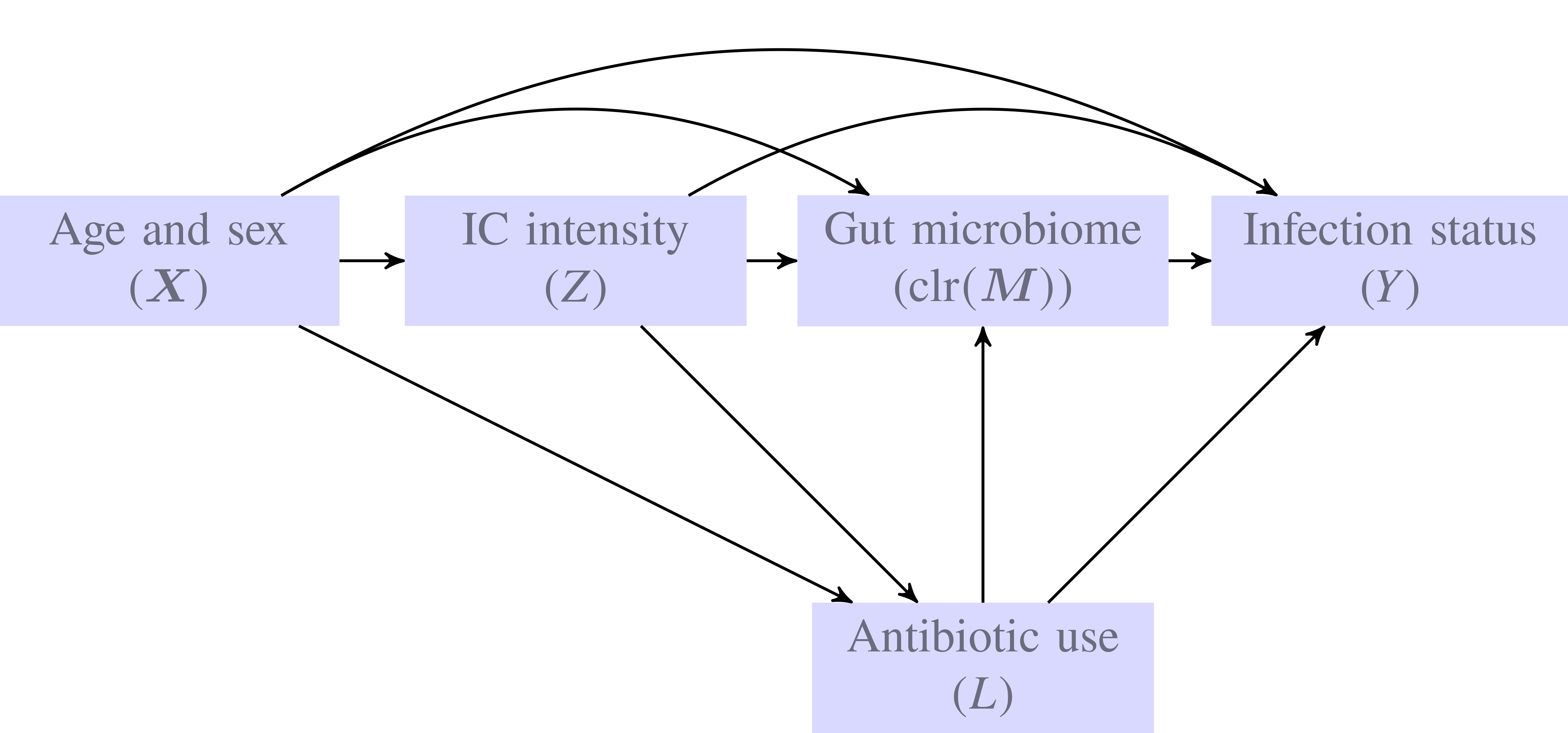}
	\end{center}
	\caption{Conceptual causal diagram for the microbiome study of acute myeloid leukemia (AML) patients.}
	\label{fig:causalmodel0}
\end{figure}

Our primary scientific question is whether gut microbiome features mediate the effect of IC intensity on infection status. As illustrated in  Fig.~\ref{fig:causalmodel0}, the exposure  is a binary indicator of IC intensity, with high-intensity regimens (e.g., fludarabine-containing and other intensive therapies) coded as 1 and low-intensity regimens (e.g., hypomethylator-based combinations and low-dose cytarabine-based therapies) coded as 0 \citep{galloway2017characterization}. The outcome of interest is  binary infection status during IC, defined using established microbiological or clinical criteria \citep{galloway2016role,galloway2017characterization}. The mediator is the gut microbiome profile, represented by genus-level abundances measured from the stool sample collected immediately prior to  infection onset or, for patients without infection, the last available sample.  Mediation analysis in this study is complicated by antibiotic use during IC, which is commonly administered for  infection prevention or treatment. Antibiotics are  known to substantially alter  gut microbiome composition \citep{schulfer2019impact} and  may also directly influence infection risk \citep{schlesinger2009infection}. We focus on broad-spectrum antibiotic use occurring  between  IC initiation and infection onset, which acts as an exposure-induced mediator--outcome confounder.
 In addition, we adjust for baseline covariates including age and sex.

\section{Methodology}
\label{sec:method}
\subsection{Preamble and notation}
\label{ss:identification}
Let $Z\in\{0,1\}$ denote a binary exposure,  $Y$ the outcome of interest, $\bm{M}=(M_1,\ldots,M_q)^\top$ a vector of  compositional mediators, $L$ an exposure-induced mediator--outcome confounder, and $\bm{X}$  a vector of baseline covariates. 
We observe an independent and identically distributed (i.i.d.) sample of size $n$,   $\{(Z_i,Y_i,\bm M_i,L_i,\bm X_i)\}_{i=1}^n$. 

Because $\bm{M}_i$ is compositional,
  we apply the centered log-ratio (clr) transformation \citep{aitchison1982statistical, lin2020analysis}, which  maps the composition from the simplex to a $(q-1)$-dimensional Euclidean subspace:
$\text{clr}(\bm{M}_i)=(\log\{M_{i1}/g(\bm{M}_i)\},\ldots, \log\{M_{iq}/g(\bm{M}_i)\})$, 
 where $g(\bm{M}_i)=(\prod_{k=1}^{q}M_{ik})^{1/q}$ denotes the geometric mean of the components of  $\bm{M}_i$.  
 The clr-transformed
components sum to zero, and each component  represents the log-ratio of a taxon's abundance to the geometric mean abundance across all taxa.
By mapping the composition to a Euclidean
subspace, the clr transformation facilitates the application of
standard statistical methods to the transformed data \citep{lin2020analysis}. We use the clr transformation because it treats all components symmetrically without requiring the specification of a reference taxon and retains a direct correspondence between transformed features and genus-level taxa, thereby facilitating  interpretation.  In the subsequent analysis, we work with
  $\text{clr}(\bm{M}_i)$ in place of $\bm{M}_i$.

Following the potential outcomes framework, let $\clr\{\bm{M}(z)\}$ denote the clr-transformed mediator value that would have been observed if $Z=z$, and let $Y\{z,\clr(\bm{m})\}$ denote the outcome value that would have been observed if $Z=z$  and $\clr(\bm{M})=\clr(\bm{m})$. The potential outcome satisfies $Y(z)=Y[z,\clr\{\bm{M}(z)\}]$. 
The total effect of $Z$ on $Y$ is  $\mathrm{TE} = E\{Y(1)\} - E\{Y(0)\}.$ Classical mediation analysis decomposes this effect into  the natural direct effect (NDE) and natural indirect effect (NIE) \citep{robins1992identifiability,pearl2001direct}:
\begin{align*}
\mathrm{NDE} =& E\left[Y\left\{1, \clr\bigl(\bm{M}(0)\bigr)\right\}\right]- E\left[Y\left\{0, \clr\bigl(\bm{M}(0)\bigr)\right\}\right], \\
\mathrm{NIE} =&E\left[Y\left\{1, \clr\bigl(\bm{M}(1)\bigr)\right\}\right]- E\left[Y\left\{1, \clr\bigl(\bm{M}(0)\bigr)\right\}\right].
\end{align*}
The NDE captures the component of the total effect not operating through the mediators, whereas the NIE represents the component mediated through $\clr(\bm M)$.
Under Supplementary Assumptions S1--S5, including the absence of exposure-induced mediator--outcome confounding (Assumption~S5), these effects are identifiable via the  mediation formula (Supplementary Proposition~S1).

\subsection{Development under exposure-induced mediator--outcome confounding}
\label{subsec:develop}
As described in Section~\ref{sec:study}, the AML microbiome study involves an exposure-induced mediator--outcome confounder $L$  (Fig.~\ref{fig:causalmodel0}). Ignoring $L$ leaves the mediator--outcome relationship confounded, whereas conditioning on $L$ as a conventional confounder is also problematic because $L$ is itself affected by the exposure. Consequently, Supplementary Assumption~S5 required for identifying the natural effects is violated, and the NDE and NIE are generally non-identifiable. To address this setting, \cite{vanderweele2014effect} introduced the 
interventional direct effect (IDE) and interventional indirect effect (IIE), defined as
\begin{align*}
\mathrm{IDE} &= E\left[Y\left\{1, \clr\bigl( \bm G(0\mid \bm X)\bigr)\right\}\right] - E\left[Y\left\{0, \clr\bigl(\bm G(0\mid \bm X)\bigr)\right\}\right], \\
\mathrm{IIE} &= E\left[Y\left\{1, \clr\bigl( \bm G(1\mid \bm X)\bigr)\right\}\right] - E\left[Y\left\{1, \clr\bigl(\bm G(0\mid \bm X)\bigr)\right\}\right],
\end{align*}
where $\bm G(z\mid \bm X)$ denotes a random draw from the conditional distribution of $\bm M$ given $\bm X$ under exposure level $z$. Unlike the natural effects, the IDE and IIE remain identifiable without Supplementary Assumption~S5; identification holds under Supplementary Assumptions~S1--S3 and~S4a (Supplementary Proposition~S2).

\subsection{Estimation of  interventional direct and indirect effects}  
\label{ss:estimation}

We develop an estimation procedure for the interventional direct and indirect effects that is tailored to the unique challenges of the AML microbiome study, including  high-dimensional, zero-inflated, and dependent microbial mediators, as well as the presence of  a binary exposure-induced mediator--outcome confounder ($L$, antibiotic use).

\cite{vanderweele2014effect} suggested estimating the IIE using the formula
\[
\label{eqn:iie}
\mathrm{IIE} =E\left\{\dfrac{Z Y}{\Pr(Z=1\mid \bm X)} \dfrac{\Pr\{\clr(\bm M)\mid Z=1, \bm X\}}{\Pr\{\clr(\bm M)\mid Z=1, L, \bm X\}}  \right\}
- E\left\{\dfrac{Z Y}{\Pr(Z=1\mid \bm X)} \dfrac{\Pr\{\clr(\bm M)\mid Z=0, \bm X\}}{\Pr\{\clr(\bm M)\mid Z=1, L, \bm X\}}  \right\}.   
\]
However, their approach requires modeling the conditional distributions $\Pr\{\clr(\bm M)\mid Z, \bm X\}$ and $\Pr\{\clr(\bm M)\mid Z,L, \bm X\}$, which is infeasible  in our AML microbiome study  because the clr-transformed microbial mediators are high-dimensional and have complex dependence structure.

Instead, leveraging the identification results in Supplementary Proposition~S2, we derive alternative identification formulas for the IDE and IIE that avoid modeling these mediator distributions. These formulas are given in Theorem~\ref{thm1} (Supplementary Proof~S3.3).

\begin{theorem}
	\label{thm1}
	Suppose that Supplementary Assumptions~S1--S3, S4a, and S6 hold. Then IDE and IIE are identified as

\[
\label{eq:IDE}
\mathrm{IDE}
=E\Biggl[\frac{(1-Z)Y}{P(z_0\mid \bm X)}
\frac{E\{Y\mid z_1, L, \clr(\bm M), \bm X\}P(L\mid z_1,\bm X)}{E\{Y\mid z_0, L, \clr(\bm M), \bm X\}
P(L\mid z_0,\clr(\bm M),\bm X)}\Biggr] -E\Biggl[\frac{(1-Z)Y}{P(z_0\mid \bm X)}
\frac{P(L\mid z_0,\bm X)}{P\{L\mid z_0,\clr(\bm M),\bm X\}}\Biggr],
\]

\[
\label{eq:IIE}
\mathrm{IIE}
=E\Biggl[\frac{ZY}{P(z_1\mid \bm X)}\frac{P(L\mid z_1,\bm X)}{
P\{L\mid z_1,\clr(\bm M),\bm X\}}\Biggr] -E\Biggl[\frac{(1-Z)Y}{P(z_0\mid \bm X)}\frac{
E\{Y\mid z_1,L,\clr(\bm M),\bm X\}P(L\mid z_1,\bm X)}{
E\{Y\mid z_0,L,\clr(\bm M),\bm X\}P\{L\mid z_0,\clr(\bm M),\bm X\}}\Biggr].
\]
\end{theorem}
\noindent where we use the shorthand  $P(z_i\mid \bm X)=\Pr(Z=i\mid \bm X)$, 
 $E\{Y\mid z_i, L, \clr(\bm M), \bm X\}=E\{Y\mid Z=i,L,\clr( \bm M),\bm X\}$, $P(L\mid z_i,\bm X)=\Pr(L\mid Z=i,\bm X)$ for $i\in\{0,1\}$, and $P\{L\mid z_0,\clr(\bm M),\bm X\}=\Pr(L\mid Z=0,\clr(\bm M),\bm X)$.

To further improve efficiency and interpretability, we restrict attention to the subset of clr-transformed mediators that are conditionally associated with the outcome given $\{Z,L,\bm X\}$. This leads to simplified expressions for the IDE and IIE, formalized in Corollary~\ref{cor:simplify} (Supplementary Proof~S3.4).
\begin{corollary}
	\label{cor:simplify}
    	Suppose that there exist subsets $\clr(\bm M^{(1)})$ and $\clr(\bm M^{(2)})$ such that $\clr(\bm M^{(1)})\cup \clr(\bm M^{(2)})=\clr(\bm M)$, $\clr(\bm M^{(1)})\cap \clr(\bm M^{(2)})=\emptyset$, $\clr(\bm M^{(1)})\nind Y\mid \{Z,L,\bm X\}$, and $\clr(\bm M^{(2)})\ind Y\mid \{Z,L,\bm X\}$.   Assume that Supplementary Assumptions~S1--S3, S4a, and S6a  hold. Then  IDE and IIE are identified as

\begin{equation}
\label{eq:IDE2}
\begin{aligned}
\mathrm{IDE}
={}&E\Biggl[
    \frac{(1-Z)Y}{P(z_0\mid\bm X)}
    \frac{
        E\{Y\mid z_1,L,\clr(\bm M^{(1)}),\bm X\}
    }{
        E\{Y\mid z_0,L,\clr(\bm M^{(1)}),\bm X\}
    }
    \frac{
        P(L\mid z_1,\bm X)
    }{
        P\{L\mid z_0,\clr(\bm M^{(1)}),\bm X\}
    }
    \Biggr]
\\[2pt]
&-
E\Biggl[
    \frac{(1-Z)Y}{P(z_0\mid\bm X)}
    \frac{
        P(L\mid z_0,\bm X)
    }{
        P\{L\mid z_0,\clr(\bm M^{(1)}),\bm X\}
    }
    \Biggr].
\end{aligned}
\end{equation}

\begin{equation}
\label{eq:IIE2}
\begin{aligned}
\mathrm{IIE}
={}&E\Biggl[\frac{ZY}{P(z_1\mid \bm X)}\frac{P(L\mid z_1,\bm X)}{
P\{L\mid z_1,\clr(\bm M^{(1)}),\bm X\}}\Biggr] \\
&-E\Biggl[\frac{(1-Z)Y}{P(z_0\mid \bm X)}\frac{
E\{Y\mid z_1,L,\clr(\bm M^{(1)}),\bm X\}P(L\mid z_1,\bm X)}{
E\{Y\mid z_0,L,\clr(\bm M^{(1)}),\bm X\}P\{L\mid z_0,\clr(\bm M^{(1)}),\bm X\}}\Biggr].
\end{aligned}
\end{equation}
where we use the shorthand  $E\{Y\mid z_i, L, \clr(\bm M^{(1)}), \bm X\}=E\{Y\mid Z=i,L,\clr(\bm M^{(1)}),\bm X\}$ for $i\in\{0,1\}$, $P\{L\mid z_0,\clr(\bm M^{(1)}),\bm X\}=\Pr(L\mid Z=0,\clr(\bm M^{(1)}),\bm X)$;  all other shorthand notation follows Theorem~\ref{thm1}.
\end{corollary}

 The IDE and IIE are then estimated based on \eqref{eq:IDE2} and \eqref{eq:IIE2}.
In the AML microbiome study, both $Z$ and $L$ are binary, so we model $\Pr(Z=1\mid \bm X)$ and $\Pr(L=1\mid Z,\bm X)$ using logistic regression with maximum likelihood estimation. 
Since $Y$ is binary and the relevant mediator subset $\clr(\bm M^{(1)})$ is unknown a priori, we  estimate $\Pr\{Y=1\mid Z,L,\clr(\bm M),\bm X\}$ using penalized logistic regression. The outcome model is constrained to include $Z$, $L$, and $\bm X$,  thereby ensuring that the estimation procedure is consistent with the identification formulas in \eqref{eq:IDE2} and \eqref{eq:IIE2}.  Let  
$\bm\beta=(\beta_0,\beta_Z,\beta_L,\bm\beta_{\clr(\bm M)}^\top,\bm\beta_{\bm X}^\top)^\top$ denote the regression coefficients in the outcome model. For the tuning parameter $\lambda$, we define
$$\hat{\bm\beta}(\lambda) = \arg\min_{\bm\beta}
\left[-\frac{2\log \{L_n(\bm\beta)\}}{n}+\sum_{k=1 }^{q}p_{\lambda}(|\beta_{M_{k}}|)\right],$$
where $L_n(\bm\beta)$ is the likelihood of the logistic regression model for $Y$,    $\beta_{M_{k}}$ is the coefficient corresponding to the $k$th clr-transformed mediator, and $p_{\lambda}(\cdot)$ denotes the smoothly clipped absolute deviation (SCAD) penalty  \citep{fan2001variable}. The tuning parameter $\lambda$ is selected by minimizing the Akaike Information Criterion (AIC) \citep{akaike1974new}:
\[
 \hat{\lambda} = \arg\min_{\lambda} \text{AIC}(\lambda)=\arg\min_{\lambda} \left[-2\log [L_n\{\hat{\bm\beta}(\lambda)\}]+2\nu(\lambda) \right],
\]
where $\nu(\lambda)$ is the number of nonzero coefficients in $\hat{\bm\beta}(\lambda)$.
The final estimator of $\bm\beta$ is chosen as $\hat{\bm\beta} = \hat{\bm\beta}(\hat{\lambda})$. The corresponding selected mediators are denoted by $\clr(\hat{\bm M}^{(1)})$.  Accordingly, the clr-transformed genus-level microbiome features selected by SCAD are regarded as candidate mediators within the joint mediation analysis.

Corollary \ref{cor:simplify} additionally requires estimation of  $\Pr\{L\mid Z, \clr(\hat{\bm M}^{(1)}),\bm X\}$. While we model $\Pr(L=1\mid Z,\bm X)$ using logistic regression, directly extending this model to include $\clr(\hat{\bm M}^{(1)})$ may lead to model incompatibility. To avoid this issue, we estimate $\Pr\{L\mid Z, \clr(\hat{\bm M}^{(1)}),\bm X\}$ using bagging with the optimal subset of deep neural networks (DNNs) \citep{mi2019bagging}, which flexibly captures nonlinear relationships and reduces overfitting. Implementation details are provided in Supplementary Note~S2.5.  

The final estimators of the IDE and IIE are obtained by plugging all component estimates into the identification formulas and computing empirical means. Supplementary Algorithm~S1 summarizes the complete procedure for estimating the IIE; estimation of the IDE proceeds analogously. We note that the specific modeling choices here are motivated by the AML microbiome application, and alternative specifications may be preferable in other settings.

\subsection{Hypothesis testing}
\label{ss:testing}

A key scientific question in the AML microbiome study is whether gut microbiome features mediate the effect of IC intensity on infection risk. This mediation effect can be quantified by the IIE  defined in Section~\ref{subsec:develop}.  Accordingly, we test $H_0: \mathrm{IIE}=0 \,\text{versus} \,  H_a: \mathrm{IIE}\neq 0$ at a prespecified significance level $\alpha$. To conduct this test, we construct a $100(1-\alpha)\%$ bootstrap  confidence interval for the IIE using the standard normal bootstrap method \citep{efron1994introduction}. The null hypothesis is rejected  if  the resulting confidence interval does not contain zero. 
This bootstrap-based testing procedure is straightforward to implement and can be efficiently accelerated using parallel computation when generating bootstrap replicates.

\section{Simulation studies}
\label{sec:simu}
We conduct simulation studies to evaluate the finite-sample performance of the proposed method and examine two NIE-based reference approaches. The simulated data are generated according to the data-generating mechanism represented by the conceptual causal diagram in Fig.~1.

Baseline covariates $\bm X = (X_1,X_2)^\top$, representing  age and sex, are sampled with replacement from the AML microbiome dataset; age is divided by 100 to place it on a scale comparable to sex. Conditional on $\bm X$, the binary exposure $Z$ and the exposure-induced mediator--outcome confounder $L$ are generated from  logistic regression models $
\Pr(Z=1\mid \bm X)=\expit(\alpha_0+\bm \alpha_{\bm X}^{\top}\bm X)
$ and 
$\Pr(L=1\mid Z,\bm X)=\expit(\gamma_{0}+\gamma_{Z}Z+\bm \gamma_{\bm X}^{\top}\bm X)$, respectively, where $\expit(x)=\exp(x)/\{1+\exp(x)\}$. To facilitate computation of the true IIE, we generate the clr-transformed mediators directly. Let $\clr(\bm M)\triangleq(\clrM_1,\ldots, \clrM_q)^\top$. We first generate $C\mid Z,L,\bm X\sim
\mathrm{Gamma}\{\eta_0,\theta(Z,L,\bm X)\}$,
where $\eta_0$ is the shape parameter and
$\theta(Z,L,\bm X)=\exp(\theta_0+\theta_{Z}Z+\theta_L L+\bm\theta_{\bm X}^{\top} \bm X)/\eta_0$
is the scale parameter. We then generate
$
\zeta
\sim
\mathrm{DiscreteUniform}
\left\{
\lfloor(q-1)a\rfloor,\ldots,\lfloor(q-1)b\rfloor
\right\}$,
where $\lfloor\cdot\rfloor$ denotes the floor function. Conditional on $C$ and $\zeta$, $\zeta$ of the first $q-1$ clr-transformed mediators are randomly selected and assigned the value $-C$, corresponding to taxa with zero read counts before zero replacement and clr transformation. The remaining $q-1-\zeta$ components are generated independently from
\[
\mathrm{Uniform}
\left(
\frac{-C(1-\zeta)}{q-1-\zeta},
\frac{(1+\zeta)C}{q-1-\zeta}
\right).
\]
The final component is set to
$\clrM_q=-\sum_{k=1}^{q-1}\clrM_{k}$
to enforce the inherent sum-to-zero constraint. This construction is chosen to approximately reproduce the zero-inflation pattern prior to clr transformation and other key characteristics of the clr-transformed microbiome data observed in the AML study. A detailed explanation of the rationale underlying the proposed data-generation mechanism is provided in Supplementary
Note~S2.6.     The binary outcome $Y$ is generated from a logistic regression model
$
\Pr(Y=1\mid Z,L,\clr(\bm M),\bm X)= \expit(\beta_0+\beta_ZZ+\beta_{L}L+\bm \beta_{\clr(\bm M)}^{\top}\clr(\bm M)+\bm \beta_{\bm X}^{\top}\bm X).
$

We set the model parameters as follows: $\alpha_0=0.2$, $\bm\alpha_{\bm X}=(-0.5,0.5)^\top$, $\gamma_{0}=0.5$, $\gamma_Z=1$, $\bm\gamma_{\bm X}=(0.5,-0.5)^\top$, $\theta_0=-1$, $\theta_Z=2$, $\theta_L=-0.2$, $\bm\theta_{\bm X}=(-0.8,-0.2)^\top$, $\eta_0=4$, $a=0.4$, $b=0.8$, $\beta_0=3$, $\beta_Z=-1.2$, $\beta_{L}=-6$,  $\bm\beta_{\bm X}=(-1.2,-1.2)^\top$, $\bm\beta_{\clr(\bm M)}=(-8,-8,0,\ldots,0)^\top$  with $q=134$, and $n=70$. Together, these parameter values define the baseline simulation setting. The primary sample size ($n=70$) and the number of clr-transformed mediators ($q=134$) are chosen to match those of the AML microbiome study. The remaining model parameters are selected with reference to summary measures observed in the AML microbiome data. Supplementary
Table~S1 reports these observed summary measures alongside their corresponding means across 500 datasets generated under the baseline simulation setting, providing a descriptive comparison between the observed and simulated data.

To investigate performance across a range of scientifically plausible scenarios, we consider additional simulation settings by varying the sample size and the parameters governing the effects along the paths $Z\rightarrow L$, $Z\rightarrow\clr(\bm M)$, and $L\rightarrow\clr(\bm M)$. Specifically, we let $\gamma_Z\in\{0,1\}$, representing the absence or presence of an effect along the path $Z\rightarrow L$. The parameters $\theta_Z\in\{0,1.5,2\}$ and $\theta_L\in\{0,-0.1,-0.2\}$ control the effects along the paths $Z\rightarrow\clr(\bm M)$ and $L\rightarrow\clr(\bm M)$, respectively. In particular, $\theta_Z$ controls whether $Z$ directly
affects the distribution of the clr-transformed mediators through $C$,
conditional on $L$ and $\bm X$. It does not parameterize the number of taxa whose
underlying absolute abundances are directly affected by the exposure; further
discussion of this distinction is provided in Supplementary Note~S2.6. The coefficient vector $\bm\beta_{\clr(\bm M)}$ is set to either $(0,\ldots,0)^\top$ or $(-8,-8,0,\ldots,0)^\top$, corresponding to scenarios in which either none or only a subset of clr-transformed mediators affect $Y$. An additional sample size of $n=200$ is also considered to assess the impact of increasing sample size.

In addition to the proposed method, we consider two NIE-based reference approaches. The first approach is constructed by adapting the proposed estimation procedure to target the NIE while ignoring  the exposure-induced mediator--outcome confounder $L$. Details are provided in  Supplementary Note~S2.7, and we refer to this approach  as the IPW-NIE method.  The second approach is the existing compositional mediation method of   \cite{sohn2022compositional}, implemented using the  R package \texttt{cmmb}. Although this method also targets the NIE, its estimation strategy differs substantially from that of the IPW-NIE method.  We refer to this approach as the Sohn-NIE method.

Results are summarized over 500 Monte Carlo replications, with 400
bootstrap replications used for each method in each simulated dataset.
Performance is assessed using  bias, standard deviation (SD),  root mean squared error (RMSE), and  the type I error rate or empirical power  at the significance level $\alpha=0.05$.  For all methods, bias and RMSE are calculated relative to the true IIE. In settings where the NIE and IIE do not coincide, these quantities for the NIE-based reference approaches represent descriptive discrepancies from the IIE rather than bias and RMSE for estimation of their  NIE estimand.
The proposed and IPW-NIE methods use standard normal
bootstrap confidence intervals, whereas the Sohn-NIE method uses the empirical bootstrap confidence interval implemented by \texttt{ci.method = "empirical"} in the \texttt{cmmb} package. Alternative bootstrap confidence intervals for the proposed and IPW-NIE methods are compared in Supplementary Note~S2.8 and Supplementary Tables~S2--S4.

Table~\ref{table:nullcase} reports results under scenarios  where $\gamma_{Z}=\theta_{Z}=0$ and/or $\bm\beta_{\clr(\bm M)}=(0,\ldots,0)^\top$. The IIE is identifiable and equals zero in all scenarios considered. When $\gamma_Z=\theta_Z=0$, the NIE is also identifiable and equals zero. In the scenarios with $\theta_Z\neq0$ considered here, the assumptions
required for identification of the NIE are violated, whereas the IIE remains identifiable and equals zero.
When $\gamma_{Z}=\theta_{Z}=0$, both the proposed method and the IPW-NIE  method maintain  type I error rates  close to the nominal level of 0.05  for both sample sizes, while the proposed method consistently achieves smaller RMSE values.  In contrast,  the Sohn-NIE method exhibits inflated  type I error rates  when  $n=70$, although its performance improves as the sample size increases. 
Across all methods,  the absolute biases are generally small relative to their corresponding SDs. When $\theta_Z\neq 0$, the proposed method continues to maintain type I error rates  close to the nominal level  for both sample sizes. However, the type I error rates of the  IPW-NIE and Sohn-NIE methods deviate substantially from the nominal level. Moreover, in these scenarios where the NIE is not identifiable, the proposed method consistently exhibits substantially smaller absolute bias than both NIE-based reference approaches.

Table~\ref{table:alternativecase} summarizes the results under scenarios with a nonzero mediation effect. In these scenarios, the assumptions
required for identification of the NIE are violated, whereas the IIE remains identifiable.
Overall, the proposed method consistently exhibits smaller absolute bias and
RMSE than both NIE-based reference approaches. 
The power of the proposed method increases with both   $\gamma_Z$ and sample size. Similar trends are observed for the IPW-NIE method. However, its power is generally lower than that of the proposed method when $n=200$. 
In contrast, the Sohn-NIE method exhibits substantially larger absolute bias and RMSE and maintains  very low power across all scenarios, even when $n=200$. Moreover, the Sohn-NIE method may encounter computational failures, resulting in missing  estimates in some Monte Carlo replications. The frequency of such failures increases markedly when $\theta_Z=2$ and is particularly severe when $n=200$.

\begin{table}[H]
	\centering
		\caption{Bias, standard deviation (SD), root mean squared error (RMSE), and type I error rate for testing $H_0: \text{mediation effect} =0$ versus $H_a: \text{mediation effect} \neq 0$ at the significance level  $\alpha=0.05$ under zero-mediation-effect scenarios. All reported  bias, SD, and RMSE values are multiplied by 100. For all methods, bias and RMSE are calculated relative to zero.}
		\label{table:nullcase}
  \small
    \setlength{\tabcolsep}{2pt}
    \renewcommand{\arraystretch}{1.1}

    \begin{adjustbox}{max width=\textwidth}
      \begin{tabular}{@{}*{21}{c}@{}}
			\toprule
			&&&&&&\multicolumn{4}{@{}c@{}}{Proposed method}&&\multicolumn{4}{@{}c@{}}{IPW-NIE method}&&\multicolumn{5}{@{}c@{}}{Sohn-NIE method}\\
			\cline{7-10} \cline{12-15} \cline{17-21}\\
			$n$&$\theta_L$&$\gamma_{Z}$&$\theta_Z$&$\bm\beta_{\clr(\bm M)}^\top$&	& Bias & SD & RMSE& Type I error rate&&Bias  & SD  & RMSE& Type I error rate&&Bias & SD & RMSE& Type I error rate&$N_{\mathrm{NA}}^\dagger$\\
			\midrule

  70 & 0 & 0 & 0 & $(0,\ldots,0)$   & &0.77 & 7.05 & 7.09 & 0.024 && 0.42 & 9.27 & 9.27 & 0.024&&-3.19 & 8.03 & 8.63 & 0.126& 0\\ 
  70 & -0.1 & 0 & 0 & $(0,\ldots,0)$  && 0.85 & 6.89 & 6.94 & 0.022 && 0.27 & 9.26 & 9.25 & 0.022&&-3.36 & 8.19 & 8.84 & 0.122& 0\\ 
 70 & -0.2 & 0 & 0 & $(0,\ldots,0)$  && 1.02 & 6.84 & 6.91 & 0.020 && 0.78 & 9.56 & 9.58 & 0.026 &&-3.56 & 8.43 & 9.14 & 0.142& 0\\ 
   70 & 0 & 0 & 0 & $(-8,-8,0,\ldots,0)$ && 1.68 & 8.96 & 9.11 & 0.020 && 0.74 & 10.16 & 10.18 & 0.026&&-1.94 & 8.62 & 8.82 & 0.102&0 \\ 
  70 & -0.1 & 0 & 0 & $(-8,-8,0,\ldots,0)$ && 1.67 & 8.45 & 8.60 & 0.024 && 1.23 & 10.38 & 10.45 & 0.036&&-1.80 & 8.73 & 8.91 & 0.100&0 \\ 
 70 & -0.2 & 0 & 0 & $(-8,-8,0,\ldots,0)$ && 1.41 & 8.37 & 8.48 & 0.026 && 1.00 & 9.78 & 9.82 & 0.034 &&-1.94 & 8.32 & 8.54 & 0.088&0\\ [1em]
70 & -0.2 & 0 & 1.5 & $(0,\ldots,0)$   && 1.68 & 7.57 & 7.74 & 0.028 && 13.36 & 8.99 & 16.10 & 0.248&&-11.77 & 6.79 & 13.59 & 0.622&0 \\ 
 70 & -0.2 & 0 & 2 & $(0,\ldots,0)$   & &0.98 & 8.09 & 8.14 & 0.030 && 12.66 & 12.63 & 17.87 & 0.184 &&-17.93 & 6.11 & 18.94 & 0.883&3\\ 
  70 & -0.2 & 1 & 1.5 & $(0,\ldots,0)$   && -2.80 & 3.85 & 4.76 & 0.026 && 7.67 & 5.59 & 9.49 & 0.110&&-16.57 & 5.40 & 17.43 & 0.928& 0\\ 
  70 & -0.2 & 1 & 2 & $(0,\ldots,0)$   && -2.29 & 4.40 & 4.95 & 0.038 && 8.65 & 5.69 & 10.35 & 0.208 &&-18.03 & 5.18 & 18.76 & 0.978&1\\ 

 \midrule

200 & 0 & 0 & 0 &  $(0,\ldots,0)$  && 0.58 & 4.60 & 4.63 & 0.046 & &1.81 & 8.64 & 8.82 & 0.032&&-1.86 & 3.58 & 4.03 & 0.056&0 \\ 
 200 & -0.1 & 0 & 0 &$(0,\ldots,0)$  && 0.67 & 4.48 & 4.53 & 0.030 && 1.53 & 9.85 & 9.96 & 0.040 &&-1.95 & 3.68 & 4.16 & 0.056&0\\ 
 200 & -0.2 & 0 & 0 & $(0,\ldots,0)$  && 0.40 & 4.43 & 4.45 & 0.034 && 1.47 & 8.76 & 8.87 & 0.028&&-2.23 & 3.74 & 4.35 & 0.068&0 \\ 
  200 & 0 & 0 & 0 &$(-8,-8,0,\ldots,0)$ && 1.00 & 5.87 & 5.95 & 0.060 && 0.49 & 10.42 & 10.42 & 0.044&&-1.22 & 3.97 & 4.15 & 0.034& 0 \\ 
 200 & -0.1 & 0 & 0 & $(-8,-8,0,\ldots,0)$ && 0.85 & 5.34 & 5.41 & 0.038 && 0.59 & 10.51 & 10.52 & 0.030&&-0.98 & 4.06 & 4.17 & 0.038&0 \\ 
 200 & -0.2 & 0 & 0 & $(-8,-8,0,\ldots,0)$ && 0.85 & 5.36 & 5.42 & 0.046 && 0.93 & 8.87 & 8.91 & 0.030&&-1.07 & 4.05 & 4.18 & 0.040&0  \\ [1em]
 200 & -0.2 & 0 & 1.5 & $(0,\ldots,0)$ && 1.33 & 6.01 & 6.14 & 0.060 && 17.50 & 5.66 & 18.39 & 0.806&&-4.73 & 3.26 & 5.74 & 0.416& 0\\ 
 200 & -0.2 & 0 & 2 & $(0,\ldots,0)$  && 1.43 & 6.25 & 6.41 & 0.060 && 15.19 & 8.74 & 17.52 & 0.626 &&-9.55 & 2.94 & 9.99 & 0.942&2\\ 
200 & -0.2 & 1 & 1.5 & $(0,\ldots,0)$  & &-1.47 & 3.48 & 3.78 & 0.086 && 8.93 & 3.63 & 9.64 & 0.832&&-6.61 & 2.13 & 6.94 & 0.956&0 \\ 
 200 & -0.2 & 1 & 2 & $(0,\ldots,0)$  && -1.22 & 3.85 & 4.03 & 0.074 && 10.03 & 3.65 & 10.68 & 0.922 &&-7.08 & 2.04 & 7.37 & 0.976&0\\

			\bottomrule			
      \end{tabular}
    \end{adjustbox}
        	\begin{tablenotes}[flushleft]
		\footnotesize
		\item[] $^\dagger$ $N_{\mathrm{NA}}$ denotes the number of Monte Carlo replications (out of 500) for which the 
		Sohn-NIE method failed because of computational issues (e.g., non-invertible matrices), resulting 
		in missing (NA) estimates. Bias, SD, RMSE, and type I error rate for this method are computed after excluding these failed replications.
	\end{tablenotes}	
\end{table}

\begin{table}[H]
	\centering
	\caption{ The true value  of the interventional indirect effect (IIE), denoted by
Truth, bias, standard deviation (SD), root mean squared error (RMSE),
		and empirical power for testing $H_0: \text{mediation effect} = 0$ versus $H_a: \text{mediation effect} \neq 0$ at the
		significance level $\alpha=0.05$ under nonzero-mediation-effect scenarios with $\theta_L=-0.2$ and $\bm\beta_{\clr(\bm M)}^\top= (-8,-8,0,\ldots,0)$. All reported  truth, bias, SD, and RMSE values are multiplied by 100. For all methods, bias and RMSE are calculated relative to the true IIE. }
	\label{table:alternativecase}
	\tabcolsep=1.4pt
	\begin{tabular*}{\textwidth}{@{\extracolsep{\fill}}cccccccccccccccccccc@{\extracolsep{\fill}}}
		\toprule
		&&&&\multicolumn{5}{@{}c@{}}{Proposed method}&&\multicolumn{4}{@{}c@{}}{IPW-NIE method}&&\multicolumn{5}{@{}c@{}}{Sohn-NIE method}\\
		\cline{5-9} \cline{11-14}  \cline{16-20}\\
		$n$&$\gamma_Z$&$\theta_Z$&&	Truth& Bias  & SD&RMSE&Power&& Bias  & SD&RMSE&Power&& Bias  & SD&RMSE&Power&$N_{\mathrm{NA}}^\dagger$\\
		\midrule

  70 & 0 & 1.5 && 15.09 & 0.35 & 11.02 & 11.01 & 0.224 && 3.11 & 13.62 & 13.96 & 0.302 &&-17.09 & 8.86 & 19.25 & 0.150 &1\\ 
  70 & 0 & 2 && 17.20 & -0.41 & 11.63 & 11.63 & 0.244 && 2.50 & 13.63 & 13.85 & 0.250&&-20.41 & 8.63 & 22.15 & 0.171 &14\\ 
  70 & 1 & 1.5 && 20.76 & -0.53 & 11.01 & 11.01 & 0.416 && -0.57 & 11.20 & 11.21 & 0.416&&-23.68 & 8.59 & 25.19 & 0.194 &0\\ 
  70 & 1 & 2 && 23.95 & -1.56 & 11.26 & 11.36 & 0.424 && -1.83 & 11.58 & 11.71 & 0.368&&-28.07 & 8.58 & 29.35 & 0.189&13 \\  
   \midrule

 200 & 0 & 1.5 && 15.09 & 1.22 & 7.09 & 7.19 & 0.630 && 6.24 & 10.51 & 12.22 &0.558 &&-14.62 & 4.11 & 15.18 & 0.037&10\\ 
 200 & 0 & 2 && 17.20 & 0.85 & 7.59 & 7.63 & 0.652 && 5.53 & 9.81 & 11.26 & 0.450 &&-17.76 & 4.31 & 18.27 & 0.055&81\\ 
 200 & 1 & 1.5 && 20.76 & -0.08 & 6.11 & 6.10 & 0.884 && 2.78 & 6.91 & 7.44 & 0.836&&-20.57 & 4.01 & 20.95 & 0.047 &8 \\ 
 200 & 1 & 2 && 23.95 & -1.25 & 6.63 & 6.74 & 0.884 && 1.31 & 7.30 & 7.41 & 0.776 &&-24.78 & 4.11 & 25.12 & 0.058&71\\ 
		
		\bottomrule			
	\end{tabular*}
	\begin{tablenotes}
		\footnotesize
		\item[] $^\dagger$ $N_{\mathrm{NA}}$ \color{black} denotes the number of Monte Carlo replications (out of 500) for which the 
		Sohn-NIE method failed because of computational issues (e.g., non-invertible matrices), resulting 
		in missing (NA) estimates. Bias, SD, RMSE, and power for this method are computed after excluding these failed replications.
	\end{tablenotes}
\end{table}

\section{Analysis of the AML microbiome data}
\label{sec:apply}

We apply the proposed method to the AML microbiome data to assess whether  gut microbiome features mediate the causal effect of  IC intensity on infection status, adjusting for the exposure-induced mediator--outcome confounder, antibiotic use, and baseline covariates age and sex (Fig.~\ref{fig:causalmodel0}; Section \ref{sec:study}). We include 70 patients with  at least one stool sample  collected between  IC initiation and either infection onset or  last follow-up.  For each patient, we use the  sample collected immediately prior to  infection onset or, for patients without infection, the last available sample. 
The mean age  was $56.2$ years (SD: 15.2), and 37 patients were female.

Following standard microbiome filtering procedures \citep{lu2019generalized},   taxa with low abundance are excluded. Specifically, we retain  genus-level taxa present in at least $10\%$ of the samples, resulting in $134$ genus-level features for  analysis.  Zero counts are replaced by the maximum rounding error of $0.5$, a common practice in the analysis of compositional and microbiome data \citep{shi2016regression}. The read counts are then converted to genus-level compositions  and  transformed using the clr transformation.

Among the 70 patients, $46$ patients received  high-intensity IC and $24$ patients received low-intensity IC. In the high-intensity group, $39$ patients used  broad-spectrum antibiotics and $15$ of them developed infections; in the low-intensity group,   $14$ patients used broad-spectrum antibiotics and $8$ of them developed infections. We estimate the average treatment effect (ATE) of IC intensity on infection status using the  Horvitz-Thompson estimator \citep{horvitz1952generalization}, with the propensity score  $\Pr(Z=1|\bm X)$ modeled via logistic regression. After adjusting for age and sex, the high-intensity IC is associated with a $23.5\%$ higher infection rate (95\% standard normal bootstrap confidence interval: $[-8.1\%,55.1\%]$). All bootstrap confidence intervals in the real-data analysis were constructed using 400 bootstrap replications.

We then apply the proposed mediation framework (Sections~\ref{ss:estimation} and~\ref{ss:testing}) to investigate whether the effect of IC intensity on infection status is mediated through gut microbiome features. For comparison, we also implement the IPW-NIE method and the Sohn-NIE method described in Section~\ref{sec:simu}, along with their corresponding approaches for estimating the NDE (the IPW-NDE method and the Sohn-NDE method).
Table \ref{table:estimate} presents the estimated effects and  95\% standard normal bootstrap confidence intervals. The proposed method indicates a significant IIE through changes in the gut microbiome at $\alpha=0.05$, whereas the estimated IDE is smaller in magnitude and is not significantly different from zero. In contrast, neither the NIE nor the NDE is significantly different from zero when using methods that ignore the exposure-induced mediator--outcome confounder, underscoring the importance of accounting for antibiotic use. The computational costs of different methods are provided in Supplementary Note~S2.9.

\begin{table}[H]
	\begin{center}
		\caption{Estimated mediation and direct effects with 95\% bootstrap confidence intervals, adjusted for  baseline covariates $\bm X$. } 
		\label{table:estimate}
        \begin{tabular*}{\columnwidth}{@{\extracolsep\fill}lcc@{\extracolsep\fill}}
			\toprule
		Effect&Estimate &  95\% confidence interval\\ 
			\midrule			 			
		Proposed method for	IIE &0.403&[0.068, 0.738]\\ 
		Proposed method for	IDE &-0.063& [-0.205, 0.079]\\[1em]
			
					IPW-NIE method &0.252&[-0.145, 0.649]\\ 
			IPW-NDE method &-0.017& [-0.352, 0.318]\\[1em]
			
							Sohn-NIE  method&0.021&[-0.248, 0.258]\\ 
			Sohn-NDE method&0.022& [-0.017, 0.065]\\
			\bottomrule
		\end{tabular*}
	\end{center}
\end{table}

The penalized logistic regression outcome model selected 11 clr-transformed genus-level microbial features  as candidate mediators (Supplementary Table~S5). Several selected features map to taxonomic groups containing members  previously reported to change during  chemotherapy-based cancer treatment or to be associated with infection or infection-related outcomes; the relevant literature is summarized in Supplementary Note~S2.10. These reported associations provide biological context but should not be interpreted as evidence of taxon-specific causal mediation effects.

To examine sensitivity to baseline covariate adjustment, which is relevant to the assumptions of no unmeasured confounding required for identification under the proposed method (Supplementary Assumptions~S2, S3, and S4a), we repeat the analysis without adjusting for the baseline covariates $\bm X$ (Supplementary Table~S6).
With adjustment for $\bm{X}$, the IIE is significantly different from zero and the IDE estimate is negative.  Without adjustment, the IIE is no longer significant and the  IDE estimate becomes positive. These differences indicate that adjustment for the measured baseline covariates materially affects the estimated effects and inferential conclusions, underscoring the importance of adequate confounding control.

We also assess the sensitivity of the results to the zero-count replacement value and the choice of log-ratio transformation. The conclusion of a significant IIE and a nonsignificant IDE remains unchanged across the
zero-count replacement values of 0.1, 0.5, and 0.9 and across the clr, additive log-ratio (alr), and isometric log-ratio (ilr) transformations. Detailed results are provided in Supplementary Note~S2.11 and Supplementary Tables~S7--S8.

\section{Discussion}
\label{sec:discuss}

In this paper, we investigate the causal relationships among IC intensity, infection status, and the on-treatment gut microbiome using data from an AML microbiome study conducted at MD Anderson. To account for antibiotic use, which is induced by IC intensity and  may confound the microbiome--infection  relationship, we adopt the interventional indirect effect  framework.  To accommodate the high dimensionality, zero inflation, and dependence of microbial mediators, we develop novel identification formulas and corresponding estimation procedures for the interventional effects. Our approach combines sparsity-inducing regularization with the clr transformation to handle compositional microbial data. 
Mediation effects are assessed using standard normal bootstrap confidence intervals. Simulation studies demonstrate that  the proposed method yields accurate effect estimates and  maintains appropriate type I error rates and power.  Application to the AML microbiome data provides evidence of a statistically significant IIE of  IC intensity on infection status through changes in the gut microbiome, with clr-transformed features corresponding to 11 genus-level taxa selected as candidate mediators. 

In the present study, we focus on microbiome measurements from  a single clinically relevant time point. 
However, the AML microbiome study collected repeated stool  samples throughout the IC treatment period. Considering all the measurements in the analysis could provide additional insight but would substantially increase the complexity of mediation analysis \citep{gilbert2025identification}. We plan to investigate this extension in our future research.  
In addition, our analysis targets the joint mediation effect through the selected microbiome features. In many applications, it is also of interest  to quantify the mediation effect of each individual mediator. This is challenging because mediators may be conditionally dependent or causally ordered, in which case some mediators can themselves act as exposure-induced mediator--outcome confounders for others \citep{zhou2022semiparametric}. Addressing these issues is nontrivial and will be explored in  future work. 

\section*{Author contributions}

Y.Z., J.W., J.S., L.W., and J.Hu. proposed the concept. J.G.-P. and S.S. provided the real data. Y.Z., J.W., J.S., L.W., and J.Hu. developed the methodology. Y.Z. and J.S. carried out the analysis. Y.Z., J.W.,  L.W., and J.Hu. drafted the initial manuscript and revised the manuscript.


\section*{Conflict of interest}
None declared.

 \section*{Funding}
This work was supported by the National Institutes of Health [P30 CA013696 to J.H., R01 AI143886 to J.H., R01 CA219896 to J.H., R01 CA256977 to J.W., P30 CA016672 to J.W.]; the Cancer Prevention and Research Institute of Texas [RP200633 to J.W.]; and the Natural Sciences and Engineering Research Council of Canada [RGPIN-2019-07052 to L.W., RGPAS-2019-00093 to L.W.,  DGECR-2019-00453 to L.W.].

\section*{Data and code availability}
The gut microbiome sequencing data supporting the findings of this article are available in the NCBI Sequence Read Archive under BioProject ID PRJNA352060 (\url{http://www.ncbi.nlm.nih.gov/sra}). Clinical metadata for the AML microbiome cohort are available through the Supplementary Materials of \cite{galloway2020gut} at \url{https://pmc.ncbi.nlm.nih.gov/articles/instance/7312220/bin/ciz777_suppl_supplementary_material.pdf}.

The R code used for the simulation studies is available at
\url{https://github.com/YX-ZHANG-1/CausalMediationMicrobiome}, and the version used to produce the results reported in this manuscript (v1.0.0) is archived on Zenodo at
\url{https://doi.org/10.5281/zenodo.21960802}.

\bibliographystyle{apalike}
\bibliography{causal}

\clearpage

\centerline{\large\bf Supplementary materials for}
\vspace{2pt}
\centerline{\large\bf ``Inverse probability weighting-based mediation analysis for microbiome data''}

\vspace{.25cm}

\setcounter{equation}{0}
\setcounter{figure}{0}
\setcounter{table}{0}
\setcounter{section}{0}

\renewcommand{\theequation}{S\arabic{equation}}
\renewcommand{\thefigure}{S\arabic{figure}}
\renewcommand{\thetable}{S\arabic{table}}
\renewcommand{\thetheorem}{S\arabic{theorem}}
\renewcommand{\theproposition}{S\arabic{proposition}}
\renewcommand{\theassumption}{S\arabic{assumption}}
\renewcommand{\thealgorithm}{S\arabic{algorithm}}
\renewcommand{\theremark}{S\arabic{remark}}
\def\thesection{S\arabic{section}}

\makeatletter
\renewcommand{\fnum@table}{Supplementary Table~\thetable}
\renewcommand{\fnum@figure}{Supplementary Figure~\thefigure}
\makeatother

\renewcommand{\refname}{Supplementary References}

\section{Supplementary Tables}
\label{s:tables}

\begin{table}[H]
\centering
\caption{Comparison of selected characteristics between the AML microbiome study and the baseline simulation setting with  $\gamma_Z=1$, $\theta_Z=2$, $\theta_L=-0.2$,
 $\bm\beta_{\clr(\bm M)}=(-8,-8,0,\ldots,0)^\top$, and $n=70$. }
	\label{table:comparedis}
	\bigskip

\begin{tabular}{lcc}
  \toprule
Characteristic& AML study&Simulation mean$^\ast$\\ 
  \midrule
  \multicolumn{3}{l}{\textit{Baseline covariates}}\\
		Mean of $X_1$&0.562&0.562 \\
		SD of $X_1$&0.152&0.151  \\
		$\Pr(X_2=1)$&0.471&0.472  \\
        
        \addlinespace
\multicolumn{3}{l}{\textit{Exposure, exposure-induced mediator--outcome confounder, and outcome}}\\
$\Pr(Z=1)$ &  0.657&0.538 \\ 
  $\Pr(L=1)$ &  0.757&0.737 \\ 
  $\Pr(Y=1)$&0.329&0.357 \\

  \addlinespace
\multicolumn{3}{l}{\textit{Microbiome characteristics}}\\
    Mean of $C$&0.959&0.815\\
Mean proportion of zero counts& 0.720&0.598  \\
25th percentile of pairwise correlations for $\clr(\bm M)$&-0.131&-0.190  \\
Median of pairwise correlations for $\clr(\bm M)$& -0.013&-0.023  \\
75th percentile of pairwise correlations for $\clr(\bm M)$& 0.119&0.163  \\
   \bottomrule
\end{tabular}

\begin{tablenotes}
		\footnotesize
		\item[] $^\ast$ Simulation means are calculated across 500 simulated
datasets generated under the baseline simulation setting.
	\end{tablenotes}	
\end{table}

\begin{table}[H]
	\centering
	\caption{ Coverage probabilities of 95\%  bootstrap confidence intervals for the mediation effect under zero-mediation-effect scenarios. Percentile CI denotes the bootstrap percentile confidence interval, BCa CI denotes the bias-corrected and accelerated bootstrap confidence interval, and Normal CI denotes the  standard normal bootstrap confidence interval.}
	\label{table:nullcase1}
	\tabcolsep=1.4pt
	\begin{tabular*}{\textwidth}{@{\extracolsep{\fill}}ccccccccccccc@{\extracolsep{\fill}}}
		\toprule
		&&&&&&\multicolumn{3}{@{}c@{}}{Proposed method}&&\multicolumn{3}{@{}c@{}}{IPW-NIE method}\\
		\cline{7-9} \cline{11-13} \\
		$n$&$\theta_L$&$\gamma_{Z}$&$\theta_Z$&$\bm\beta_{\clr(\bm M)}^\top$&	& Percentile CI & BCa CI & Normal CI &&Percentile CI & BCa CI & Normal CI \\
		\midrule

		70 & 0 & 0 & 0 & $(0,\ldots,0)$   & & 0.992 & 0.986 & 0.976 && 0.998 & 0.996 & 0.976 \\ 
		70 & -0.1 & 0 & 0 & $(0,\ldots,0)$  && 0.990 & 0.984 & 0.978 && 1.000 & 0.998 & 0.978 \\ 
		70 & -0.2 & 0 & 0 & $(0,\ldots,0)$  &&  0.990 & 0.986 & 0.980 && 0.998 & 0.994 & 0.974 \\ 
        		70 & 0 & 0 & 0 & $(-8,-8,0,\ldots,0)$ && 0.988 & 0.986 & 0.980 && 0.998 & 0.998 & 0.974 \\ 
		70 & -0.1 & 0 & 0 & $(-8,-8,0,\ldots,0)$ &&0.992 & 0.984 & 0.976 && 0.996 & 0.998 & 0.964 \\
		70 & -0.2 & 0 & 0 & $(-8,-8,0,\ldots,0)$ &&  0.996 & 0.990 & 0.974 && 0.998 & 0.998 & 0.966 \\[1em]
		70 & -0.2 & 0 & 1.5 & $(0,\ldots,0)$   && 0.996 & 0.986 & 0.972 && 0.914 & 0.906 & 0.752 \\ 
		70 & -0.2 & 0 & 2 & $(0,\ldots,0)$   & &0.994 & 0.980 & 0.970 && 0.950 & 0.940 & 0.816 \\ 
		70 & -0.2 & 1 & 1.5 & $(0,\ldots,0)$   && 0.952 & 0.934 & 0.974 && 0.910 & 0.866 & 0.890 \\ 
		70 & -0.2 & 1 & 2 & $(0,\ldots,0)$   &&0.962 & 0.940 & 0.962 && 0.846 & 0.778 & 0.792 \\ 

	\midrule
        
		200 & 0 & 0 & 0 &  $(0,\ldots,0)$  && 0.998 & 0.998 & 0.954 && 1.000 & 1.000 & 0.968 \\ 
		200 & -0.1 & 0 & 0 &$(0,\ldots,0)$  &&  0.998 & 0.998 & 0.970 && 0.996 & 0.998 & 0.960 \\ 
		200 & -0.2 & 0 & 0 & $(0,\ldots,0)$  && 0.998 & 0.998 & 0.966 && 1.000 & 1.000 & 0.972 \\ 
        		200 & 0 & 0 & 0 &$(-8,-8,0,\ldots,0)$ &&  0.988 & 0.986 & 0.940 && 1.000 & 1.000 & 0.956 \\ 
		200 & -0.1 & 0 & 0 & $(-8,-8,0,\ldots,0)$ && 0.988 & 0.988 & 0.962 && 1.000 & 1.000 & 0.970 \\ 
		200 & -0.2 & 0 & 0 & $(-8,-8,0,\ldots,0)$ && 0.996 & 0.994 & 0.954 && 1.000 & 1.000 & 0.970 \\ [1em]
		200 & -0.2 & 0 & 1.5 & $(0,\ldots,0)$ &&  0.998 & 1.000 & 0.940 && 0.248 & 0.364 & 0.194 \\ 
		200 & -0.2 & 0 & 2 & $(0,\ldots,0)$  && 0.994 & 0.994 & 0.940 && 0.390 & 0.508 & 0.374 \\ 
		200 & -0.2 & 1 & 1.5 & $(0,\ldots,0)$  & &  0.972 & 0.970 & 0.914 && 0.152 & 0.180 & 0.168 \\
		200 & -0.2 & 1 & 2 & $(0,\ldots,0)$  && 0.970 & 0.964 & 0.926 && 0.036 & 0.042 & 0.078 \\

		\bottomrule			
	\end{tabular*}
\end{table}

\begin{table}[H]
	\centering
	\caption{The true value  of the interventional indirect effect (IIE), denoted by
Truth, and coverage probabilities of 95\%  bootstrap confidence intervals for the mediation effect under nonzero-mediation-effect scenarios with $\theta_L=-0.2$ and $\bm\beta_{\clr(\bm M)}^\top= (-8,-8,0,\ldots,0)$. Percentile CI denotes the bootstrap percentile confidence interval, BCa CI denotes the bias-corrected and accelerated bootstrap confidence interval, and Normal CI denotes the  standard normal bootstrap confidence interval. All reported  truth values are multiplied by 100.}
	\label{table:alternativecase1}
	\tabcolsep=1.4pt
	\begin{tabular*}{\textwidth}{@{\extracolsep{\fill}}cccccccccccc@{\extracolsep{\fill}}}
		\toprule
		&&&&\multicolumn{4}{@{}c@{}}{Proposed method}&&\multicolumn{3}{@{}c@{}}{IPW-NIE method}\\
		\cline{5-8} \cline{10-12} \\
	$n$ & $\gamma_Z$ & $\theta_Z$ &	&Truth& Percentile CI & BCa CI & Normal CI &&Percentile CI & BCa CI & Normal CI \\
		\midrule

 70 & 0 & 1.5 && 15.09 & 0.990 & 0.988 & 0.958 && 1.000 & 1.000 & 0.966 \\ 
 70 & 0 & 2 && 17.20 & 0.990 & 0.984 & 0.964 && 1.000 & 1.000 & 0.976 \\ 
 70 & 1 & 1.5 && 20.76 & 0.964 & 0.954 & 0.950 && 0.996 & 0.994 & 0.984 \\ 
 70 & 1 & 2 && 23.95 & 0.970 & 0.968 & 0.954 && 0.998 & 0.992 & 0.974 \\
\midrule

 200 & 0 & 1.5 && 15.09 & 0.986 & 0.986 & 0.960 && 0.952 & 0.966 & 0.892 \\ 
 200 & 0 & 2 && 17.20 & 0.986 & 0.988 & 0.944 && 0.990 & 0.992 & 0.954 \\ 
 200 & 1 & 1.5 && 20.76 & 0.988 & 0.986 & 0.968 && 0.984 & 0.988 & 0.964 \\ 
 200 & 1 & 2 && 23.95 & 0.978 & 0.986 & 0.958 && 1.000 & 0.998 & 0.986 \\
		
		\bottomrule		
	\end{tabular*}
\end{table}

\begin{table}[H]
	\centering
	\caption{ The true value  of the interventional indirect effect (IIE), denoted by
Truth,  and empirical power for testing $H_0: \text{mediation effect} = 0$ versus $H_a: \text{mediation effect} \neq 0$ at the
		significance level  $\alpha=0.05$ under nonzero-mediation-effect scenarios with $\theta_L=-0.2$ and $\beta_{\mathrm{clr}(M)}^\top= (-8,-8,0,\ldots,0)$. Percentile CI denotes the bootstrap percentile confidence interval, BCa CI denotes the bias-corrected and accelerated bootstrap confidence interval, and Normal CI denotes the  standard normal bootstrap confidence interval. All reported truth values are multiplied by 100.}
	\label{table:alternativecase2}
	\tabcolsep=1.4pt
	\begin{tabular*}{\textwidth}{@{\extracolsep{\fill}}cccccccccccc@{\extracolsep{\fill}}}
		\toprule
		&&&&\multicolumn{4}{@{}c@{}}{Proposed method}&&\multicolumn{3}{@{}c@{}}{IPW-NIE method}\\
		\cline{5-8} \cline{10-12} \\
		$n$ & $\gamma_Z$ & $\theta_Z$ &	&Truth& Percentile CI & BCa CI & Normal CI &&Percentile CI & BCa CI & Normal CI \\
		\midrule

		70 & 0 & 1.5 && 15.09 & 0.126 & 0.120 & 0.224 && 0.118 & 0.102 & 0.302 \\  
		70 & 0 & 2 && 17.20 & 0.124 & 0.142 & 0.244 && 0.128 & 0.140 & 0.250 \\ 
		70 & 1 & 1.5 && 20.76 & 0.322 & 0.344 &0.416 && 0.204 & 0.176 & 0.416 \\ 
	 70 & 1 & 2 && 23.95 & 0.358 & 0.360 & 0.424 && 0.200 & 0.188 & 0.368 \\
		\midrule
		
	200 & 0 & 1.5 && 15.09 & 0.564 & 0.578 & 0.630 && 0.586 & 0.510 & 0.558 \\ 
	200 & 0 & 2 && 17.20 & 0.664 & 0.686 & 0.652 && 0.512 & 0.448 & 0.450 \\ 
	 200 & 1 & 1.5 && 20.76 & 0.904 & 0.896 & 0.884 && 0.874 & 0.826 & 0.836 \\ 
	200 & 1 & 2 && 23.95 & 0.940 & 0.944 & 0.884 && 0.850 & 0.796 & 0.776 \\
		
		\bottomrule		
	\end{tabular*}
\end{table}

\begin{table}[H]
\centering
\setlength{\tabcolsep}{1pt}
\begin{threeparttable}
\caption{Details of the selected genus-level microbial features in the
AML microbiome data analysis.}
\label{table:selectedgenera}

\begin{tabular*}{\columnwidth}
{@{\extracolsep{\fill}}cccc@{\extracolsep{\fill}}}
\toprule
OTU & Genus & Family & Kingdom \\
\midrule
GFQLact9  & \textit{Lactococcus}
           & Streptococcaceae & Bacteria \\
GG7The26  & \textit{Bacteroides}
           & Bacteroidaceae & Bacteria \\
GWTAmn51  & \textit{Enterobacter}
           & Enterobacteriaceae & Bacteria \\
Unc03xpn  & \textit{Lachnoclostridium}
           & Lachnospiraceae & Bacteria \\
Unc04n2w  & Actinomycetaceae (Unc04n2w)
           & Actinomycetaceae & Bacteria \\
Unc05m27  & Ruminococcaceae (Unc05m27)
           & Ruminococcaceae & Bacteria \\
Unc05mrd  & \textit{Akkermansia}
           & Verrucomicrobiaceae & Bacteria \\
Unc36622  & Lachnospiraceae (Unc36622)
           & Lachnospiraceae & Bacteria \\
Unc94755  & \textit{Alistipes}
           & Rikenellaceae & Bacteria \\
UncCl291  & Clostridiales (UncCl291)
           & Clostridiales (UncCl291) & Bacteria \\
UncM5880  & \textit{Methanobrevibacter}
           & Methanobacteriaceae & Archaea \\
\bottomrule
\end{tabular*}

\begin{tablenotes}[flushleft]
\footnotesize
\item[]  \textit{Note:} Taxonomic annotations are reproduced from the original taxon
table. For features not classified at the genus level, the available
higher-level taxonomic annotation is followed by the original feature
identifier in parentheses. ``Clostridiales (UncCl291)'' is retained as
the family-level annotation assigned in the original data.
\end{tablenotes}
\end{threeparttable}
\end{table}

\begin{table}[H]
	\begin{center}
		\caption{Estimated interventional indirect and direct effects (IIE and IDE) with  95\% bootstrap confidence intervals,  unadjusted for baseline covariates $\bm X$.}  
		\label{table:estimatenox}
	\begin{tabular*}{\columnwidth}{@{\extracolsep\fill}lcc@{\extracolsep\fill}}
			\toprule
			Effect&Estimate&  95\% confidence interval\\ 
			\midrule			 			
		Proposed method for	IIE &0.067&[-0.183, 0.318]\\ 
		Proposed method for	IDE &0.023& [-0.251, 0.296]\\
			\bottomrule
		\end{tabular*}
	\end{center}
\end{table}

\begin{table}[H]
	\begin{center}
		\caption{Estimated interventional indirect and direct effects (IIE and IDE) with 95\% bootstrap confidence intervals under different zero-count replacement values, adjusted for baseline covariates $\bm X$.}
		\label{table:zerosensitivity}
		\begin{tabular*}{\columnwidth}{@{\extracolsep\fill}cccc@{\extracolsep\fill}}
			\toprule
			Zero-count replacement value & Effect & Estimate & 95\% confidence interval\\
			\midrule
			\multirow{2}{*}{0.1} 
			& Proposed method for IIE & 0.443 & [0.102, 0.784]\\
			& Proposed method for IDE & -0.115 & [-0.262, 0.031]\\
			\addlinespace
			\multirow{2}{*}{0.5} 
			& Proposed method for IIE & 0.403 & [0.068, 0.738]\\
			& Proposed method for IDE & -0.063 & [-0.205, 0.079]\\
			\addlinespace
			\multirow{2}{*}{0.9} 
			& Proposed method for IIE & 0.442 & [0.111, 0.773]\\
			& Proposed method for IDE & -0.109 & [-0.249, 0.031]\\
			\bottomrule
		\end{tabular*}
	\end{center}
\end{table}

\begin{table}[H]
	\begin{center}
\caption{Estimated interventional indirect and direct effects (IIE and IDE) with 95\% bootstrap confidence intervals under different log-ratio transformations of $\bm{M}$, adjusted for baseline covariates $\bm X$. The clr transformation corresponds to the transformation used in the proposed method. For the additive log-ratio (alr) transformation, the last genus-level taxon, unide146, is used as the reference component. For the isometric log-ratio (ilr) transformation, the clr-transformed composition is projected onto the basic (Helmert) orthonormal basis implemented by \texttt{ilrBase()} with \texttt{method = "basic"} in the R package \texttt{compositions}, yielding $q-1$ ilr coordinates.}
\label{table:transformationsensitivity}
		\begin{tabular*}{\columnwidth}{@{\extracolsep\fill}cccc@{\extracolsep\fill}}
			\toprule
			Transformation of $\bm{M}$& Effect & Estimate & 95\% confidence interval\\ 
			\midrule
			\multirow{2}{*}{clr} & IIE  & 0.403&[0.068, 0.738]\\
			    & IDE  & -0.063& [-0.205, 0.079]\\
			\addlinespace
				\multirow{2}{*}{alr} & IIE  & 0.403&[0.049, 0.756]\\
			    & IDE  & -0.023& [-0.170, 0.125]\\
			\addlinespace
				\multirow{2}{*}{ilr} &  IIE  &0.400 & [0.061, 0.740]\\
			    & IDE  & -0.067 & [-0.211, 0.077]\\
			\bottomrule
		\end{tabular*}
	\end{center}
\end{table}

\section{Supplementary Notes}
\label{s:notes}

\subsection{Regularity conditions }
\label{s:conditions}
\begin{assumption}[Consistency]
	\label{assump:1}
	$\clr(\bm{M})=\clr\{\bm{M}(z)\}$ when $Z=z$; $Y=Y\{z,\clr(\bm{m})\}$ when $Z=z$ and $\clr(\bm{M})=\clr(\bm{m})$.
\end{assumption} 

	\begin{assumption}[No unmeasured $Z-Y$ confounding]
		\label{assump:2}
		For all $z, \bm m$, $Z \ind Y\{z,\clr(\bm m)\}\mid \bm X;$
	\end{assumption}
    
	\begin{assumption}[No unmeasured $Z-\bm M$ confounding]
		\label{assump:3}
	For all $z$, $Z\ind \clr\{\bm M(z)\}\mid \bm X;$
	\end{assumption}
    
	\begin{assumption}[No unmeasured $\bm M-Y$ confounding]
		\label{assump:4}
		For all $z, \bm m$, $\clr(\bm M) \ind Y\{z,\clr( \bm m)\}\mid \{Z,\bm X\};$
	\end{assumption}
    
	\begin{assumption}[No exposure-induced $\bm M-Y$ confounding]
		\label{assump:5}
		For all $\bm m$,  $\clr\{\bm M(0)\}\ind Y\{1,\clr(\bm m)\}\mid \bm X$.
	\end{assumption}

\setcounter{assumption}{3}
\renewcommand{\theassumption}{S4\alph{assumption}}
\setcounter{assumption}{0}

    \begin{assumption}[No unmeasured $\bm M-Y$ confounding]
		\label{assump:6}
		For all $z, \bm m, Y\{z,\clr( \bm m)\}\ind \clr(\bm M)\mid \{Z,L,\bm X\}$.
	\end{assumption}

\setcounter{assumption}{5} 
\renewcommand{\theassumption}{S\arabic{assumption}}

    \begin{assumption}[Positivity]
		\label{assump:7}
	For $z=0,1$ and all $\bm x$, $\Pr(Z=z\mid \bm X=\bm x)>0$; for all $l$, $\bm m$, and $\bm x$, $E\{Y\mid Z=0, L=l, \clr(\bm M)=\clr(\bm m),\bm X=\bm x\}>0$;  and for $z=0,1$ and all $l$, $\bm m$, and $\bm x$, $\Pr\{L=l\mid Z=z, \clr(\bm M)=\clr(\bm m), \bm X=\bm x\}>0$.
	\end{assumption}

    \setcounter{assumption}{5}
\renewcommand{\theassumption}{S6\alph{assumption}}
\setcounter{assumption}{0}
	\begin{assumption}[Positivity]
		\label{assump:8}
	For $z=0,1$ and all $\bm x$, $\Pr(Z=z\mid \bm X=\bm x)>0$; for all $l$, $\clr(\bm m^{(1)})$, and $\bm x$, $E\{Y\mid Z=0, L=l, \clr(\bm M^{(1)})=\clr(\bm m^{(1)}),\bm X=\bm x\}>0$; and for $z=0,1$ and all $l$, $\clr(\bm m^{(1)})$, and $\bm x$, $\Pr\{L=l\mid Z=z,\clr(\bm M^{(1)})=\clr(\bm m^{(1)}),\bm X=\bm x\}>0$.
	\end{assumption}

\subsection{Identification of natural direct and indirect effects}
\begin{proposition}\citep[Mediation formula,][]{pearl2001direct}
	\label{prop1}
	Suppose that Assumptions \ref{assump:1}--\ref{assump:5} hold, 
	then the NDE and NIE are identifiable. If  $\bm X$ and $\clr(\bm M)$ are discrete, then
	\begin{flalign*}
	\mathrm{NDE} &= \sum\limits_{\clr(\bm m),\bm x} \left[ E\left\{Y\mid z_1, \clr( \bm m), \bm x\right\} - E\left\{Y\mid z_0, \clr( \bm m), \bm x\right\} \right]  P\left\{\clr(\bm m)\mid z_0, \bm x\right\}P(\bm x), \\
	\mathrm{NIE} &=\sum\limits_{\clr(\bm m), \bm x}  E\left\{Y\mid z_1, \clr(\bm m), \bm x\right\} \left[P\left\{\clr(\bm m)\mid z_1, \bm x\right\} - P\left\{\clr(\bm m)\mid z_0, \bm x\right\}\right]P(\bm x),
	\end{flalign*}
	where, following the standard shorthand in the mediation literature, $E\{Y\mid z_1, \clr(\bm m),\bm x\} = E\{Y\mid Z=1, \clr(\bm M)=\clr(\bm m), \bm X=\bm x\}$, $E\{Y\mid z_0, \clr(\bm m),\bm x\} = E\{Y\mid Z=0, \clr( \bm M)=\clr( \bm m), \bm X=\bm x\}$, $P\{\clr(\bm m)\mid z_1, \bm x\} = \Pr\{\clr(\bm M)=\clr(\bm m)\mid Z=1,\bm X=\bm x\}$, $P\{\clr(\bm m)\mid z_0, \bm x\} = \Pr\{\clr(\bm M)=\clr(\bm m)\mid Z=0,\bm X=\bm x\}$, and $P(\bm x) =\Pr(\bm X=\bm x)$.
\end{proposition}

The proof of Proposition~\ref{prop1} is provided in Section \ref{s:proofprop1}.  Under the following nonparametric structural equation models (NPSEM): 
\begin{equation}
\label{eqn:npsem}
\bm X = f_{\bm X} (\bm \epsilon_{\bm X}),\, Z(\bm x) = f_Z (\bm x,\epsilon_Z),\, \clr \{\bm M(\bm x,z)\} = f_{\clr(\bm M)}(\bm x,z,\bm \epsilon_{\bm M}), \,
 Y\{\bm x,z,\clr(\bm m)\} = f_Y\{\bm x,z,\clr(\bm m), \epsilon_Y\}.
\end{equation}
Assumptions \ref{assump:2}--\ref{assump:5} follow from the independent error (IE) assumption that $\bm \epsilon_{\bm X} \ind \epsilon_Z \ind \bm \epsilon_{\bm M} \ind \epsilon_Y$.  Supplementary Figure~\ref{fig:causalmodel1}  shows the causal diagram corresponding to the NPSEM in  \eqref{eqn:npsem}.

\begin{figure}[!htbp]
	\begin{center}
		\begin{tikzpicture}[->,>=stealth',shorten >=1pt,auto,node distance=2.3cm,
		semithick, scale=0.50,
		pre/.style={-,>=stealth,semithick,blue,line width = 1pt}]
		\tikzstyle{every state}=[fill=none,draw=black,text=black]
		\node[state] (A)                    {$Z$};
		\node[state] (M) [right of=A] {$\clr(\bm M)$};
		\node[state]         (Y) [right of=M] {$Y$};
		\node[state]         (X) [left of=A] {$\bm X$};
		\path   (A) edge node {} (M)
		(M) edge node {} (Y)
		(A) edge  [bend left] node  {} (Y)
		(X) edge node {} (A)
		(X) edge  [bend left] node {} (M)
		(X) edge  [bend left] node {} (Y);
		\end{tikzpicture}
	\end{center}
	\caption{A causal diagram corresponding to  the nonparametric structural equation models (NPSEM) in \eqref{eqn:npsem}.}
	\label{fig:causalmodel1}
\end{figure}

\subsection{Identification of interventional direct and indirect effects}

\begin{proposition}\citep{vanderweele2014effect}
	\label{prop:vanderweele}
	Suppose that  Assumptions \ref{assump:1} -- \ref{assump:3} and \ref{assump:6} hold.	
	Then  IDE and IIE are identifiable. If $\bm X$, $L$, and $\clr(\bm M)$ are discrete, then
	\begin{equation}
	\label{eqn:identify1}
	\begin{aligned}
	\mathrm{IDE}
	&= \sum\limits_{l,\clr(\bm m),\bm x} \left[E\{Y\mid z_1, l, \clr(\bm m), \bm x\} P(l\mid z_1, \bm x) - E\{Y\mid z_0, l, \clr(\bm m), \bm x\} P(l\mid z_0, \bm x)\right]
     P\{\clr(\bm m)\mid z_0, \bm x\} P(\bm x); \\
	\mathrm{IIE}
	&=\sum\limits_{l,\clr(\bm m),\bm x} E\{Y\mid z_1, l, \clr(\bm m), \bm x\} P(l\mid z_1, \bm x) \left[P\{\clr(\bm m)\mid z_1, \bm x\} - P\{\clr(\bm m)\mid z_0, \bm x\}\right] P(\bm x), 
	\end{aligned}
	\end{equation}
where we use the shorthand that $E\{Y\mid z_1, l, \clr(\bm m), \bm x\}=E\{Y\mid Z=1,L=l,\clr( \bm M)=\clr(\bm m),\bm X=\bm x\}$, $E\{Y\mid z_0, l, \clr(\bm m), \bm x\}=E\{Y\mid Z=0,L=l,\clr(\bm M)=\clr(\bm m),\bm X=\bm x\}$, $P(l\mid z_1,\bm x)=\Pr(L=l\mid Z=1,\bm X=\bm x)$, $P(l\mid z_0,\bm x)=\Pr(L=l\mid Z=0,\bm X=\bm x)$, and other shorthand notation follows Proposition~\ref{prop1}.
\end{proposition}

The proof of Proposition~\ref{prop:vanderweele} is provided in Section \ref{s:proofprop2}. The identification expression in~\eqref{eqn:identify1} naturally extends to continuous $\bm X$, $L$, and $\clr(\bm M)$ by replacing summation with  integration. Assumptions~\ref{assump:2}, \ref{assump:3}, and \ref{assump:6} are satisfied under the causal diagram in Figure~1 of the main manuscript; they would also hold if the causal association between $L$ and $Y$ were confounded by additional unmeasured factors.  

\begin{remark}
	Assumption \ref{assump:5} is   a ``cross-world'' independence assumption \citep{robins2010alternative}, meaning that it cannot be validated by any randomized experiment involving the variables in Figure \ref{fig:causalmodel1}.
	In contrast, all the assumptions required for Proposition \ref{prop:vanderweele}  are ``single-world'' and can  be guaranteed under randomization of both $Z$ and $\clr(\bm M)$.
\end{remark}

\begin{remark}
	If $L$ is absent, then the identification formulas for the IDE and IIE reduce to those for the NDE and NIE.
\end{remark}

\subsection{Algorithm for estimating the IIE}
\begin{algorithm}[H]
	\caption{Proposed inverse probability weighting approach for estimating the $\mathrm{IIE}$}
	\label{algorithm}
	\begin{algorithmic}
		\item [1.] Fit logistic regression models for $\Pr(Z=1\mid \bm X; \bm\alpha)$ and $\Pr(L=1\mid Z,\bm X;\bm \gamma)$ using the maximum likelihood estimation. 	Let $\hat{\Pr}(Z=z\mid \bm X) = {\Pr}(Z=z\mid \bm X; \hat{\bm \alpha})$ and  
$\hat{\Pr}(L\mid Z=z,\bm X)
=
\Pr(L\mid Z=z,\bm X;\hat{\bm\gamma}),
z\in\{0,1\}$, where $\hat{\bm\alpha}$ and $\hat{\bm\gamma}$ denote the  maximum likelihood estimates of $\bm\alpha$ and $\bm \gamma$, respectively.
		\item [2.] Estimate $E(Y\mid Z=z, L, \clr(\bm M^{(1)}), \bm X)$, $z=0,1$, using the  penalized logistic regression procedure described earlier.   Let $\clr(\hat{\bm M}^{(1)})$ denote the selected set of mediators and $\hat{E}(Y\mid Z=z, L, \clr(\hat{\bm M}^{(1)}), \bm X)$ the corresponding fitted conditional means.
		\item [3.] Estimate $\Pr(L\mid Z=z, \clr(\hat{\bm M}^{(1)}), \bm X), z=0,1$, using bagging with the optimal subset of DNNs.  Let $\hat{\Pr}(L\mid Z=z,\clr(\hat{\bm M}^{(1)}),\bm X)$ denote the corresponding fitted conditional probabilities.
		\item [4.] The estimator of the IIE is
		\begin{multline*}
		\hat{\mathrm{IIE}} =  \mathbb{P}_n \left\{\dfrac{Z Y}{\hat{\Pr}(Z=1\mid \bm X)} \dfrac{\hat{\Pr}(L\mid Z=1, \bm X)}{\hat{\Pr}(L\mid Z=1,  \clr(\hat{\bm M}^{(1)}), \bm X)}\right\}\\
		- \mathbb{P}_n\left\{\dfrac{(1-Z) Y}{\hat{\Pr}(Z=0\mid \bm X)} \dfrac{\hat{E}(Y\mid Z=1, L,  \clr(\hat{\bm M}^{(1)}), \bm X) \hat{\Pr}(L\mid Z=1, \bm  X)}{\hat{E}(Y\mid Z=0, L,  \clr(\hat{\bm M}^{(1)}), \bm X) \hat{\Pr}(L\mid Z=0,  \clr(\hat{\bm M}^{(1)}), \bm X)}  \right\},
		\end{multline*}
		where $\mathbb{P}_n$ denotes the empirical mean operator.	
	\end{algorithmic}
\end{algorithm}

\subsection{The implementation details of  bagging with the optimal subset of deep neural networks}

To implement bagging with the optimal subset of deep neural networks (DNNs), 
we follow the approach of \cite{mi2019bagging} and use their  R package  \texttt{deepTL}. Documentation for  \texttt{deepTL}  is available at \url{https://github.com/SkadiEye/deepTL}.

In both the simulation studies and real data analysis, we use fully connected feedforward DNNs with one hidden layer containing five hidden nodes. The activation function is the rectified linear unit (ReLU), defined as $\text{ReLU}(t)= \max(0,t)$.
The tuning parameter $\lambda$ for the $L_1$ penalty function is set to $10^{-5}$.
Mini-batch stochastic gradient descent is performed using a batch size of 20 for a maximum of 100 training epochs, with Adam used as the adaptive optimization algorithm \citep{kingma2014adam}. We train an ensemble of 20 DNNs, and the final ensemble is automatically selected according to the criterion proposed by \cite{mi2019bagging}. The simulation results are found to be robust to reasonable choices of these tuning parameters.

\subsection{Rationale underlying the proposed data-generation mechanism for the clr-transformed mediators}

Recall that, in Section~4 of the main manuscript, we generate the clr-transformed mediators directly rather than simulating microbiome read counts. This strategy substantially reduces the computational burden of evaluating the true interventional indirect effect while retaining selected empirical characteristics of the observed AML microbiome data relevant to the simulation design. Here we describe the rationale underlying the proposed data-generation mechanism.

In the AML microbiome study, microbiome measurements are recorded as read counts. Let $R_{ik}$ denote the observed read count of the $k$th taxon for the $i$th sample, where $i\in\{1,\ldots,n\}$ and $k\in\{1,\ldots,q\}$. Let
\[
M_{ik}
=
\frac{R_{ik}}{\sum_{j=1}^{q}R_{ij}}
\]
denote the corresponding compositional relative abundance. Thus, each relative abundance is defined relative to the total read count across all taxa.

It is useful to distinguish exposure effects on the underlying absolute-abundance scale from those on the compositional relative-abundance and clr-transformed scales. From a biological perspective, an exposure may directly affect the underlying absolute abundances of only a subset of microbial taxa. However, such sparsity does not generally carry over to the compositional relative-abundance scale. Because each relative abundance is defined with respect to the total abundance across all taxa, changes in a subset of taxa can change the common denominator and consequently alter the relative abundances of taxa whose underlying absolute abundances are not directly affected.

The clr transformation further expresses each taxon's relative abundance as the log-ratio of that relative abundance to the geometric mean of the relative abundances across all taxa:
\[
\clr(M_{ik})
=
\log\left\{\frac{M_{ik}}{g(\bm M_i)}\right\},
\qquad
g(\bm M_i)
=
\left(\prod_{j=1}^{q}M_{ij}\right)^{1/q}.
\]
Thus, the clr-transformed value of taxon $k$ depends not only on its own relative abundance $M_{ik}$, but also on the relative abundances of all other taxa through the common geometric mean $g(\bm M_i)$. Consequently, when the underlying absolute abundances of a subset of taxa change, the resulting changes in the compositional relative abundances can alter $g(\bm M_i)$, which in turn can change the clr-transformed values corresponding to taxa whose absolute abundances are not directly affected. Therefore, sparsity on the underlying absolute-abundance scale does not generally translate into sparsity on the clr-transformed scale.

We next describe how the empirical characteristics of the AML microbiome data motivate our direct generation of the clr-transformed mediators. Because microbiome read counts are highly zero-inflated, zero counts are replaced by a small positive constant $\delta$ (e.g., $\delta=0.5$) before calculating the relative abundances and applying the clr transformation. Let
\[
\widetilde R_{ik}
=
\begin{cases}
R_{ik}, & R_{ik}>0,\\
\delta, & R_{ik}=0,
\end{cases}
\qquad
\widetilde T_i
=
\sum_{k=1}^{q}\widetilde R_{ik},
\]
and define
\[
\widetilde M_{ik}
=
\frac{\widetilde R_{ik}}{\widetilde T_i}.
\]
For a taxon with an original zero read count, $\widetilde R_{ik}=\delta$, and its clr-transformed value is
\[
\clr\left(\frac{\delta}{\widetilde T_i}\right)
=
\log\left(\frac{\delta}{\widetilde T_i}\right)
-
\frac{1}{q}
\sum_{k=1}^{q}
\log\left(\frac{\widetilde R_{ik}}{\widetilde T_i}\right)
\triangleq B_i.
\]
Since $\widetilde R_{ik}\geq\delta$ for all taxa and $\widetilde R_{ik}>\delta$ for at least one taxon, we have $B_i<0$. Define $C_i=-B_i$, so that $C_i>0$.

\begin{figure}[H]
\centering
\includegraphics[width=0.8\textwidth]{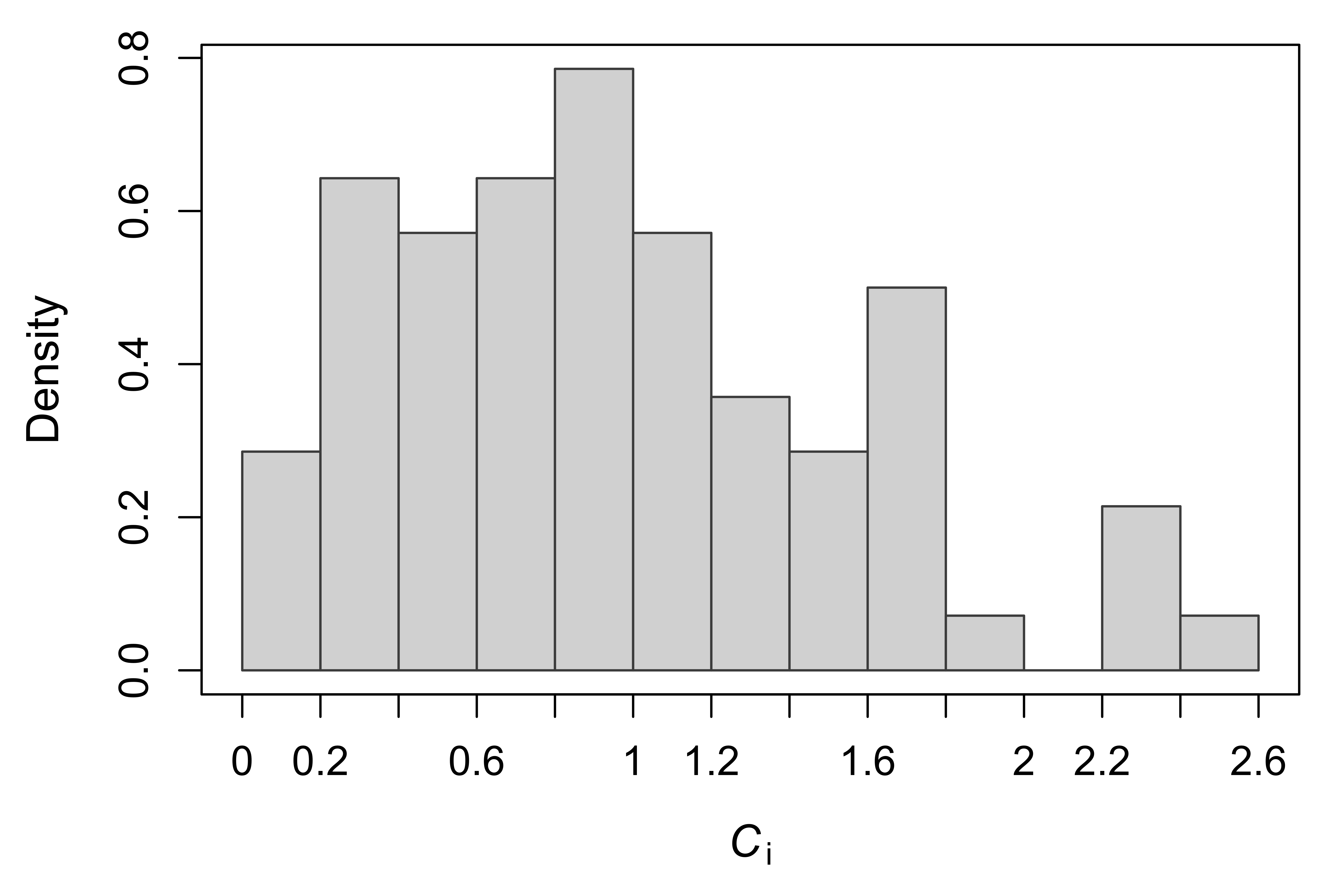}
\caption{Histogram of the observed values of $C_i$ in the AML microbiome study plotted on the density scale.}
\label{fig:gammafit}
\end{figure}

Supplementary Figure~\ref{fig:gammafit} shows that the observed values of $C_i$ in the AML microbiome study are positive and right-skewed. Motivated by this empirical feature, we assume
\[
C_i\mid(Z_i,L_i,\bm X_i)
\sim
\mathrm{Gamma}
\left\{
\eta_0,
\theta(Z_i,L_i,\bm X_i)
\right\},
\]
where $\eta_0$ is the shape parameter and
\[
\theta(Z_i,L_i,\bm X_i)
=
\frac{
\exp\left(
\theta_0+\theta_ZZ_i+\theta_LL_i+
\bm\theta_{\bm X}^{\top}\bm X_i
\right)
}{\eta_0}
\]
is the scale parameter. Under this parameterization,
\[
E(C_i\mid Z_i,L_i,\bm X_i)
=
\exp\left(
\theta_0+\theta_ZZ_i+\theta_LL_i+
\bm\theta_{\bm X}^{\top}\bm X_i
\right),
\]
and
\[
\operatorname{Var}(C_i\mid Z_i,L_i,\bm X_i)
=
\frac{
\{E(C_i\mid Z_i,L_i,\bm X_i)\}^2
}{\eta_0},
\]
corresponding to a Gamma generalized linear model with a log link.

Under this data-generation mechanism, $\theta_Z$ controls whether $Z$ directly affects the distribution of the clr-transformed mediators through $C_i$, conditional on $L_i$ and $\bm X_i$. It does not parameterize how many taxa are directly affected by the exposure on the underlying absolute-abundance scale.

To evaluate the Gamma generalized linear modeling assumption for $C_i$, we fit the corresponding Gamma generalized linear model to the AML data and assess its adequacy using randomized quantile residual diagnostics implemented in the R package \texttt{DHARMa}. The Kolmogorov--Smirnov test ($p$-value $=0.55$), dispersion test ($p$-value $=0.60$), and outlier test ($p$-value $=1.00$) provide no evidence of lack of fit, overdispersion, or excessive outliers, supporting the use of the Gamma generalized linear model for the conditional distribution of $C_i$.

For sample $i$, let $\zeta_i$ denote the number of taxa among the first $q-1$ components that have zero read counts before zero replacement and therefore take the clr-transformed value $B_i=-C_i$. Examination of the AML microbiome data indicates that the distribution of $\zeta_i$ is reasonably approximated by a discrete uniform distribution. We therefore generate
\[
\zeta_i
\sim
\mathrm{DiscreteUniform}
\left\{
\lfloor(q-1)a\rfloor,
\ldots,
\lfloor(q-1)b\rfloor
\right\},
\]
where $\lfloor\cdot\rfloor$ denotes the floor function. The constants $a$ and $b$ determine the range of zero-count proportions in the simulated microbiome data. In our simulation studies, we set $a=0.4$ and $b=0.8$, allowing the proportion of zero read counts to range from moderate to high levels while remaining representative of the zero inflation observed in the AML microbiome study.

Conditional on $\zeta_i$, exactly $\zeta_i$ of the first $q-1$ clr-transformed mediators take the value $B_i=-C_i$. The remaining $q-1-\zeta_i$ components are assumed, for simplicity, to follow a uniform distribution conditional on $C_i$ and $\zeta_i$. Denote these  components by
\[
\left\{
\clrM_{i(1)},\ldots,
\clrM_{i(q-1-\zeta_i)}
\right\}.
\]
Because clr-transformed components satisfy the sum-to-zero constraint,
\[
\clrM_{iq}
=
-\sum_{k=1}^{q-1}\clrM_{ik}.
\]
In addition, $B_i=-C_i$ is the clr-transformed value corresponding to a zero read count, so we require
\[
\clrM_{iq}\geq B_i.
\]
Since $\zeta_i$ of the first $q-1$ components equal $B_i$, this requirement implies
\[
-\left\{
\zeta_iB_i+
\sum_{s=1}^{q-1-\zeta_i}\clrM_{i(s)}
\right\}
\geq B_i,
\]
or equivalently,
\[
\sum_{s=1}^{q-1-\zeta_i}
\clrM_{i(s)}
\leq
-(1+\zeta_i)B_i
=
(1+\zeta_i)C_i.
\]
A sufficient condition for this constraint is
\[
\clrM_{i(s)}
\leq
\frac{(1+\zeta_i)C_i}
{q-1-\zeta_i},
\qquad
s=1,\ldots,q-1-\zeta_i,
\]
which determines the upper endpoint of the uniform distribution.

To center the marginal distribution of each of the first $q-1$ clr-transformed mediators, we choose the lower endpoint $a_i$ such that
\[
E(\clrM_{ik}\mid C_i,\zeta_i)=0,
\qquad
k=1,\ldots,q-1.
\]
This yields
\[
a_i
=
-\frac{C_i(1-\zeta_i)}
{q-1-\zeta_i}.
\]
Under random assignment of the $\zeta_i$ zero components among the first $q-1$ taxa, the marginal distribution of each of these clr-transformed mediators is therefore
\[
\clrM_{ik}\mid C_i,\zeta_i
\sim
\frac{\zeta_i}{q-1}
\mathbb{I}(\clrM_{ik}=-C_i)
+
\left(
1-\frac{\zeta_i}{q-1}
\right)
\mathrm{Uniform}
\left(
a_i,
\frac{(1+\zeta_i)C_i}{q-1-\zeta_i}
\right),
\qquad
k=1,\ldots,q-1.
\]
It is straightforward to verify that
\[
a_i\geq -C_i
\]
for $q\geq2$, and
\[
a_i
<
\frac{(1+\zeta_i)C_i}
{q-1-\zeta_i},
\]
so the support of the uniform distribution is well defined. Finally, setting
\[
\clrM_{iq}
=
-\sum_{k=1}^{q-1}\clrM_{ik}
\]
ensures that the simulated clr-transformed mediator vector satisfies the inherent sum-to-zero constraint, and $E(\clrM_{iq})=0$.

\subsection{Estimation and hypothesis testing methods for the natural effects}
\label{sec:NIE}

Following the same idea as the proposed estimation method for the interventional effects, the natural direct effect (NDE) and natural indirect effect (NIE) can be estimated using identification formulas that avoid modeling the conditional distributions of $\clr(\bm M)$. The result is given in Proposition~\ref{thm:NIE}.

\begin{proposition}
	\label{thm:NIE}
	Suppose that there exist subsets $\clr(\bm M^{(1)})$ and $\clr(\bm M^{(2)})$ such that $\clr(\bm M^{(1)})\cup \clr(\bm M^{(2)})=\clr(\bm M)$, $\clr(\bm M^{(1)})\cap \clr(\bm M^{(2)})=\emptyset$, $\clr(\bm M^{(1)})\nind Y\mid \{Z,\bm X\}$, and $\clr(\bm M^{(2)})\ind Y\mid \{Z,\bm X\}$. Under Assumptions~\ref{assump:1}--\ref{assump:5}, NDE and NIE are identified as
	\begin{equation*}
	\mathrm{NDE} = E\left\{\dfrac{(1-Z) Y}{\Pr(Z=0\mid \bm X)} \dfrac{E(Y\mid Z=1,  \clr(\bm M^{(1)}), \bm X)}{E(Y\mid Z=0, \clr(\bm M^{(1)}), \bm X)}  \right\}
	- E\left\{\dfrac{(1-Z) Y}{\Pr(Z=0\mid \bm X)}\right\},
	\label{eq:NDE2}
\end{equation*}
	
\begin{equation*}
	\mathrm{NIE} = E\left\{\dfrac{Z Y}{\Pr(Z=1\mid \bm X)}\right\}
	- E\left\{\dfrac{(1-Z) Y}{\Pr(Z=0\mid \bm X)} \dfrac{E(Y\mid Z=1,  \clr(\bm M^{(1)}), \bm X)}{E(Y\mid Z=0,  \clr(\bm M^{(1)}), \bm X)}  \right\}.
	\label{eq:NIE2}
	\end{equation*}
	\end{proposition}

Based on Proposition~\ref{thm:NIE}, estimation of the NDE and NIE proceeds analogously to the procedure in Section~3.3 in the main manuscript, with two key differences:  
(i) modeling of the exposure-induced mediator--outcome confounder $L$ is not required; and  
(ii) we estimate $E(Y\mid Z=z,\clr(\bm M^{(1)}),\bm X)$ for $z=0,1$ instead of the conditional expectations involving $L$.  
These conditional expectations can be obtained using penalized logistic regression, following the same strategy described in Section~3.3 in the main manuscript but omitting $L$ from the outcome model.
Hypothesis testing for the mediation effect can be performed by testing  
\[
H_0: \mathrm{NIE}=0 \quad \text{versus} \quad H_a: \mathrm{NIE}\neq 0,
\]
using the same bootstrap confidence interval approach described in Section~3.4 in the main manuscript for the IIE.

We refer to the estimation and testing procedures for the NDE as the IPW-NDE method, and the corresponding procedures for the NIE as the IPW-NIE method.

\subsection{Comparison of three bootstrap confidence intervals in the simulation studies}
\label{sec:bsci}

Note that the type-I error rates and power reported for the proposed method and the IPW-NIE method in Tables~1 and 2 of the main manuscript are based on the standard normal bootstrap confidence interval (denoted as Normal CI), following the procedure described in Section~3.4 of the main manuscript and Supplementary Note~\ref{sec:NIE}. To compare the performance of different bootstrap confidence intervals, we additionally consider the bootstrap percentile confidence interval (denoted as Percentile CI; \cite{efron1994introduction}) and the bias-corrected and accelerated bootstrap confidence interval (denoted as BCa CI; \cite{efron1987better}). The coverage probabilities of the resulting 95\% confidence intervals are presented in Supplementary Tables~\ref{table:nullcase1} and \ref{table:alternativecase1}, and the corresponding power results under nonzero mediation effect scenarios are reported in Supplementary Table~\ref{table:alternativecase2}.

Under zero mediation effect scenarios (Supplementary Table~\ref{table:nullcase1}), the type-I error rate is equal to one minus the coverage probability. For the proposed method, the coverage probabilities of the Normal CI are generally close to the nominal level of 0.95, implying type-I error rates close to the nominal significance level of 0.05. The Percentile CI and BCa CI often exhibit higher coverage probabilities than the Normal CI and, in several scenarios, yield coverage probabilities close to one, indicating more conservative inference and correspondingly lower type-I error rates. Similar patterns are observed for the IPW-NIE method when the NIE is identifiable ($\gamma_{Z}=\theta_{Z}=0$). However, when the NIE is not identifiable ($\theta_Z\neq 0$), all three confidence intervals for the IPW-NIE method exhibit substantial departures from the nominal level, regardless of the bootstrap procedure used.

Under nonzero mediation effect scenarios (Supplementary Tables~\ref{table:alternativecase1} and \ref{table:alternativecase2}), the Normal CI for the proposed method generally achieves coverage probabilities close to the nominal level while often providing higher power than the Percentile CI and BCa CI, particularly when the sample size is small ($n=70$). As the sample size increases to $n=200$, the power differences among the three confidence interval procedures become less pronounced, although the Percentile CI and BCa CI often exhibit  slightly higher coverage probabilities than the Normal CI. Similar patterns are observed for the IPW-NIE method. In general, the coverage probabilities of the Normal CI for the IPW-NIE method exhibit larger departures from the nominal level than those of the proposed method. 

Overall, the Normal CI used by the proposed method achieves coverage probabilities closest to the nominal level across the scenarios considered while maintaining appropriate type-I error control and often providing higher power than the more conservative Percentile CI and BCa CI. Taken together, these results suggest that the Normal CI provides a favorable balance between coverage accuracy and statistical power for the proposed method.

\subsection{Computational cost for the AML microbiome data analysis}
\label{sec:real}

Regarding computational cost, the real-data analysis was performed on a 40-core node equipped with an  Intel Cascade Lake processor, with all 40  cores used for parallel computing. All reported wall-clock times include the bootstrap calculations used to construct the confidence intervals.
For the analysis reported in Table~3 of the main manuscript, the proposed method required 3.45 seconds to obtain the IIE and IDE estimates and their corresponding confidence intervals. The IPW-NIE and IPW-NDE methods required 2.19 seconds, whereas the Sohn-NIE and Sohn-NDE methods required substantially longer, with a total wall-clock time of 1,968.07 seconds.

\subsection{Biological context for the selected candidate mediators}

\label{sec:candidatemed}

The penalized logistic regression outcome model selected 11 clr-transformed genus-level microbial features as candidate mediators (Supplementary Table~S5). Several selected features map to taxonomic groups containing members  previously reported to change during  chemotherapy-based cancer treatment, including Streptococcaceae, Bacteroidaceae, Enterobacteriaceae, Lachnospiraceae, Actinomycetaceae, Ruminococcaceae, Verrucomicrobiaceae, Rikenellaceae,  Clostridiales, and  Methanobacteriaceae \citep{damico2021gut,papanicolas2021conventional,terrisse2021intestinal,zhang2021intestinal,wu2022effect,xu2024gut}. For example, \citet{xu2024gut} reported that the abundance of Enterobacteriaceae decreased significantly after R-CHOP-based treatment in patients with diffuse large B-cell lymphoma. In a longitudinal study of patients with breast cancer, \citet{wu2022effect} reported chemotherapy-regimen-associated abundance  changes involving \textit{Actinomyces} (family Actinomycetaceae), Rikenellaceae, \textit{Akkermansia} (family Verrucomicrobiaceae), and \textit{Methanobrevibacter} (family Methanobacteriaceae).

Several of these taxonomic groups have additionally been associated with infection or infection-related outcomes \citep{vincent2013reductions,khanna2016gut,sangster2016bacterial,hakim2018gut,vazquez2023characterization,mcmahon2025utilization}. For example, \citet{vincent2013reductions} found that lower relative abundances of Bacteroidaceae and a family-level group within Clostridiales  in the gut microbiome were associated with subsequent  nosocomial \textit{Clostridioides difficile} infection. These previously reported associations provide biological context for the selected features but should not be interpreted as independent confirmation of taxon-specific causal mediation effects.

\subsection{Sensitivity to zero-count replacement and compositional data
transformation}
\label{sec:sensitivity}
To assess  sensitivity  to zero-count replacement before the clr transformation, we repeat the proposed mediation analysis using replacement values of 0.1, 0.5, and 0.9, where 0.5 is used in the primary analysis. These values allow us to examine the sensitivity of the results over a range of replacement values below the minimum nonzero count of 1 observed in the AML microbiome data. As shown in Supplementary Table~S7, the estimated IIEs range from 0.403 to 0.443, and all corresponding 95\% bootstrap confidence intervals exclude zero. The estimated IDEs range from $-0.115$ to $-0.063$, with all confidence intervals including zero. These results indicate that both the effect estimates and inferential conclusions remain stable across the replacement values examined.

We also assess sensitivity to the compositional data transformation by repeating the analysis using the additive log-ratio (alr) and isometric log-ratio (ilr) transformations in addition to the primary clr transformation \citep{lin2020analysis}. As shown in Supplementary Table~S8, the estimated IIEs are 0.403, 0.403, and 0.400 under the clr, alr, and ilr transformations, respectively, and all corresponding 95\% bootstrap confidence intervals exclude zero. The IDE confidence intervals include zero under all three transformations. Thus, the conclusion of a significant IIE but a nonsignificant IDE remains unchanged across the log-ratio transformations considered.

\section{Supplementary Proofs}
\label{s:proofs}

\subsection{Proof of Proposition \ref{prop1}}
\label{s:proofprop1}

\begin{proof}
If Assumptions \ref{assump:1}--\ref{assump:5}  hold, then
\begin{small}
\begin{flalign*}
{\rm NDE} 
=& E\left[Y\left\{1, \clr\left(\bm{M}(0)\right)\right\}\right]- E\left[Y\left\{0, \clr\left(\bm{M}(0)\right)\right\}\right]\\
=&\sum_{ \clr(\bm m), \bm x}E[Y\{1, \clr(\bm m)\}\mid \clr \{\bm M(0)\}=\clr(\bm m),\bm X=\bm x]\Pr[\clr \{\bm M(0)\}=\clr(\bm m)\mid \bm X=\bm x]\Pr(\bm X=\bm x)\\
&-\sum_{\clr(\bm m), \bm x}E[Y\{0, \clr(\bm m)\}\mid \clr\{\bm M(0)\}=\clr(\bm m),\bm X=\bm x]\Pr[\clr\{\bm M(0)\}=\clr(\bm m)\mid \bm X=\bm x]\Pr(\bm X=\bm x)\\
=&\sum_{\clr(\bm m), \bm x}E[Y\{1, \clr(\bm m)\}\mid \bm X=\bm x]\Pr[\clr\{\bm M(0)\}=\clr(\bm m)\mid \bm X=\bm x]\Pr(\bm X=\bm x)\\
&-\sum_{\clr(\bm m), \bm x}E[Y\{0, \clr(\bm m)\}\mid \bm X=\bm x]\Pr[\clr\{\bm M(0)\}=\clr(\bm m)\mid \bm X=\bm x]\Pr(\bm X=\bm x)\\
=&\sum_{\clr(\bm m), \bm x}E[Y\{1, \clr(\bm m)\}\mid Z=1,\bm X=\bm x]\Pr[\clr\{\bm M(0)\}=\clr(\bm m)\mid Z=0,\bm X=\bm x]\Pr(\bm X=\bm x)\\
&-\sum_{\clr(\bm m), \bm x}E[Y\{0, \clr(\bm m)\}\mid Z=0,\bm X=\bm x]\Pr[\clr\{\bm M(0)\}=\clr(\bm m)\mid Z=0,\bm X=\bm x]\Pr(\bm X=\bm x)\\
=&\sum_{\clr(\bm m), \bm x}E[Y\{1, \clr(\bm m)\}\mid Z=1,\clr(\bm M)=\clr(\bm m),\bm X=\bm x]\Pr[\clr\{\bm M(0)\}=\clr(\bm m)\mid Z=0,\bm X=\bm x]\Pr(\bm X=\bm x)\\
-&\sum_{\clr(\bm m), \bm x}E[Y\{0, \clr(\bm m)\}\mid Z=0,\clr(\bm M)=\clr(\bm m),\bm X=\bm x]\Pr[\clr\{\bm M(0)\}=\clr(\bm m)\mid Z=0,\bm X=\bm x]\Pr(\bm X=\bm x)\\
=&\sum_{\clr(\bm m), \bm x}E\{Y\mid Z=1,\clr(\bm M)=\clr(\bm m),\bm X=\bm x\}\Pr\{\clr(\bm M)=\clr(\bm m)\mid Z=0,\bm X=\bm x\}\Pr(\bm X=\bm x)\\
&-\sum_{\clr(\bm m), \bm x}E\{Y\mid Z=0,\clr(\bm M)=\clr(\bm m),\bm X=\bm x\}\Pr\{\clr(\bm M)=\clr(\bm m)\mid Z=0,\bm X=\bm x\}\Pr(\bm X=\bm x)\\
=&\sum_{\clr(\bm m), \bm x}\Big[\left\{E\left(Y\mid Z=1,\clr(\bm M)=\clr(\bm m),\bm X=\bm x\right)-E\left(Y\mid Z=0,\clr(\bm M)=\clr(\bm m),\bm X=\bm x\right)\right\}\\
& \qquad\times \Pr\{\clr(\bm M)=\clr(\bm m)\mid Z=0,\bm X=\bm x\}\Pr(\bm X=\bm x)\Big].
\end{flalign*}
\end{small}

\begin{small}
\begin{flalign*}
{\rm NIE}
=&E\left[Y\left\{1, \clr\left(\bm{M}(1)\right)\right\}\right]- E\left[Y\left\{1, \clr\left(\bm{M}(0)\right)\right\}\right]\\
=&\sum_{\clr(\bm m), \bm x}E[Y\{1, \clr(\bm m)\}\mid \clr \{\bm M(1)\}=\clr(\bm m),\bm X=\bm x]
\Pr[\clr\{\bm M(1)\}=\clr(\bm m)\mid \bm X=\bm x]\Pr(\bm X=\bm x)\\
&-\sum_{\clr(\bm m), \bm x}E[Y\{1, \clr(\bm m)\}\mid \clr\{\bm M(0)\}=\clr(\bm m),\bm X=\bm x]\Pr[\clr\{\bm M(0)\}=\clr(\bm m)\mid \bm X=\bm x]\Pr(\bm X=\bm x)\\
=&\sum_{\clr(\bm m), \bm x}E[Y\{1, \clr(\bm m)\}\mid \bm X=\bm x]\Pr[\clr \{\bm M(1)\}=\clr(\bm m)\mid \bm X=\bm x]\Pr(\bm X=\bm x)\\
&-\sum_{\clr(\bm m), \bm x}E[Y\{1, \clr(\bm m)\}\mid \bm X=\bm x]\Pr[\clr\{\bm M(0)\}=\clr(\bm m)\mid \bm X=\bm x]\Pr(\bm X=\bm x)\\
=&\sum_{\clr(\bm m), \bm x}E[Y\{1, \clr(\bm m)\}\mid Z=1,\bm X=\bm x]\Pr[\clr \{\bm M(1)\}=\clr(\bm m)\mid Z=1,\bm X=\bm x]\Pr(\bm X=\bm x)\\
&-\sum_{\clr(\bm m), \bm x}E[Y\{1, \clr(\bm m)\}\mid Z=1,\bm X=\bm x]\Pr[\clr\{\bm M(0)\}=\clr(\bm m)\mid Z=0,\bm X=\bm x]\Pr(\bm X=\bm x)\\
=&\sum_{\clr(\bm m), \bm x}E[Y\{1, \clr(\bm m)\}\mid Z=1,\clr(\bm M)=\clr(\bm m),\bm X=\bm x]\Pr[\clr \{\bm M(1)\}=\clr(\bm m)\mid Z=1,\bm X=\bm x]\Pr(\bm X=\bm x)\\
-&\sum_{\clr(\bm m), \bm x}E[Y\{1, \clr(\bm m)\}\mid Z=1,\clr(\bm M)=\clr(\bm m),\bm X=\bm x]\Pr[\clr\{\bm M(0)\}=\clr(\bm m)\mid Z=0,\bm X=\bm x]\Pr(\bm X=\bm x)\\
=&\sum_{\clr(\bm m), \bm x}E\{Y\mid Z=1,\clr(\bm M)=\clr(\bm m),\bm X=\bm x\}\Pr\{\clr(\bm M)=\clr(\bm m)\mid Z=1,\bm X=\bm x\}\Pr(\bm X=\bm x)\\
&-\sum_{\clr(\bm m), \bm x}E\{Y\mid Z=1,\clr(\bm M)=\clr(\bm m),\bm X=\bm x\}\Pr\{\clr(\bm M)=\clr(\bm m)\mid Z=0,\bm X=\bm x\}\Pr(\bm X=\bm x)\\
=&\sum_{\clr(\bm m), \bm x}\Big[E\left\{Y\mid Z=1,\clr(\bm M)=\clr(\bm m),\bm X=\bm x\right\}\\
&\qquad \times \left\{\Pr\left(\clr(\bm M)=\clr(\bm m)\mid Z=1,\bm X=\bm x\right)-\Pr\left(\clr(\bm M)=\clr(\bm m)\mid Z=0,\bm X=\bm x\right)\right\} \Pr(\bm X=\bm x)\Big].
\end{flalign*}
\end{small}
\end{proof}

\subsection{Proof of Proposition \ref{prop:vanderweele}}
\label{s:proofprop2}
\begin{proof}
If Assumptions \ref{assump:1}--\ref{assump:3} and \ref{assump:6}  hold, then
\begin{small}
\begin{flalign*}
{\rm IDE} 
=&E\left[Y\left\{1, \clr\left( \bm G(0\mid \bm X)\right)\right\}\right] - E\left[Y\left\{0, \clr\left(\bm G(0\mid \bm X)\right)\right\}\right]\\
=&\sum_{\clr(\bm m), \bm x}E[Y\{1, \clr(\bm m)\}\mid \clr\{\bm G(0\mid \bm x)\}=\clr(\bm m),\bm X=\bm x]\Pr[\clr\{\bm G(0\mid \bm x)\}=\clr(\bm m)\mid \bm X=\bm x]\Pr(\bm X=\bm x)\\
&-\sum_{\clr(\bm m), \bm x}E[Y\{0, \clr(\bm m)\}\mid \clr\{\bm G(0\mid \bm x)\}=\clr(\bm m),\bm X=\bm x]\Pr[\clr\{\bm G(0\mid \bm x)\}=\clr(\bm m)\mid \bm X=\bm x]\Pr(\bm X=\bm x)\\
=&\sum_{\clr(\bm m), \bm x}E[Y\{1, \clr(\bm m)\}\mid \bm X=\bm x]\Pr[\clr \{\bm M(0)\}=\clr(\bm m)\mid \bm X=\bm x]\Pr(\bm X=\bm x)\\
&-\sum_{\clr(\bm m), \bm x}E[Y\{0, \clr(\bm m)\}\mid \bm X=\bm x]\Pr[\clr\{\bm M(0)\}=\clr(\bm m)\mid \bm X=\bm x]\Pr(\bm X=\bm x)\\
=&\sum_{\clr(\bm m), \bm x}E[Y\{1, \clr(\bm m)\}\mid Z=1,\bm X=\bm x]\Pr[\clr\{\bm M(0)\}=\clr(\bm m)\mid Z=0,\bm X=\bm x]\Pr(\bm X=\bm x)\\
&-\sum_{\clr(\bm m), \bm x}E[Y\{0, \clr(\bm m)\}\mid Z=0,\bm X=\bm x]\Pr[\clr\{\bm M(0)\}=\clr(\bm m)\mid Z=0,\bm X=\bm x]\Pr(\bm X=\bm x)\\
=&\sum_{l,\clr(\bm m), \bm x}\Big[E\left\{Y\{1, \clr(\bm m)\}\mid Z=1,L=l,\bm X=\bm x\right\}\Pr(L=l\mid Z=1,\bm X=\bm x)\\
&\qquad\times \Pr[\clr\{\bm M(0)\}=\clr(\bm m)\mid Z=0,\bm X=\bm x]\Pr(\bm X=\bm x)\Big]\\
&-\sum_{l,\clr(\bm m), \bm x}\Big[E\left\{Y\{0, \clr(\bm m)\}\mid Z=0,L=l,\bm X=\bm x\right\}\Pr(L=l\mid Z=0,\bm X=\bm x)\\
&\qquad \times \Pr[\clr \{\bm M(0)\}=\clr(\bm m)\mid Z=0,\bm X=\bm x]\Pr(\bm X=\bm x)\Big]\\
=&\sum_{l,\clr(\bm m), \bm x}\Big[E\left\{Y\{1, \clr(\bm m)\}\mid Z=1,L=l,\clr(\bm M)=\clr(\bm m),\bm X=\bm x\right\}\Pr(L=l\mid Z=1,\bm X=\bm x)\\
&\qquad\times \Pr[\clr\{\bm M(0)\}=\clr(\bm m)\mid Z=0,\bm X=\bm x]\Pr(\bm X=\bm x)\Big]\\
&-\sum_{l,\clr(\bm m), \bm x}\Big[E\left\{Y\{0, \clr(\bm m)\}\mid Z=0,L=l,\clr(\bm M)=\clr(\bm m),\bm X=\bm x\right\}\Pr(L=l\mid Z=0,\bm X=\bm x)\\
&\qquad\times \Pr[\clr\{\bm M(0)\}=\clr(\bm m)\mid Z=0,\bm X=\bm x]\Pr(\bm X=\bm x)\Big]\\
=&\sum_{l,\clr(\bm m), \bm x}\Big[E\{Y\mid Z=1,L=l,\clr(\bm M)=\clr(\bm m),\bm X=\bm x\}\Pr(L=l\mid Z=1,\bm X=\bm x)\\
&\qquad\times \Pr\{\clr(\bm M)=\clr(\bm m)\mid Z=0,\bm X=\bm x\}\Pr(\bm X=\bm x)\Big]\\
&-\sum_{l,\clr(\bm m), \bm x}\Big[E\{Y\mid Z=0,L=l,\clr(\bm M)=\clr(\bm m),\bm X=\bm x\}\Pr(L=l\mid Z=0,\bm X=\bm x)\\
&\qquad\times\Pr\{\clr(\bm M)=\clr(\bm m)\mid Z=0,\bm X=\bm x\}\Pr(\bm X=\bm x)\Big]\\
=&\sum_{l,\clr(\bm m), \bm x}\Big[\big\{E(Y\mid Z=1,L=l,\clr(\bm M)=\clr(\bm m),\bm X=\bm x)\Pr(L=l\mid Z=1,\bm X=\bm x)\\
&\quad-E\{Y\mid Z=0,L=l,\clr(\bm M)=\clr(\bm m),\bm X=\bm x\}\Pr(L=l\mid Z=0,\bm X=\bm x)\big\}\\
 &\qquad\times \Pr\{\clr(\bm M)=\clr(\bm m)\mid Z=0,\bm X=\bm x\}\Pr(\bm X=\bm x)\Big].
\end{flalign*}
\end{small}

\begin{small}
\begin{flalign*}
{\rm IIE} 
=&E\left[Y\left\{1, \clr\left( \bm G(1\mid \bm X)\right)\right\}\right] - E\left[Y\left\{1, \clr\left(\bm G(0\mid \bm X)\right)\right\}\right]\\
=&\sum_{\clr(\bm m), \bm x}E[Y\{1,\clr(\bm m)\}\mid \clr\{\bm G(1\mid x)\}=\clr(\bm m),\bm X=\bm x]\Pr[\clr\{\bm G(1\mid \bm x)\}=\clr(\bm m)\mid \bm X=\bm x]\Pr(\bm X=\bm x)\\
&-\sum_{\clr(\bm m), \bm x}E[Y\{1, \clr(\bm m)\}\mid \clr\{\bm G(0\mid \bm x)\}=\clr(\bm m),\bm X=\bm x]\Pr[\clr\{\bm G(0\mid \bm x)\}=\clr(\bm m)\mid \bm X=\bm x]\Pr(\bm X=\bm x)\\
=&\sum_{\clr(\bm m), \bm x}E[Y\{1, \clr(\bm m)\}\mid \bm X=\bm x]\Pr[\clr \{\bm M(1)\}=\clr(\bm m)\mid \bm X=\bm x]\Pr(\bm X=\bm x)\\
&-\sum_{\clr(\bm m), \bm x}E[Y\{1, \clr(\bm m)\}\mid \bm X=\bm x]\Pr[\clr\{\bm M(0)\}=\clr(\bm m)\mid \bm X=\bm x]\Pr(\bm X=\bm x)\\
=&\sum_{\clr(\bm m), \bm x}E[Y\left\{1, \clr(\bm m)\right\}\mid Z=1,\bm X=\bm x]\Pr[\clr \{\bm M(1)\}=\clr(\bm m)\mid Z=1,\bm X=\bm x]\Pr(\bm X=\bm x)\\
&-\sum_{\clr(\bm m), \bm x}E[Y\{1, \clr(\bm m)\}\mid Z=1,\bm X=\bm x]\Pr[\clr\{\bm M(0)\}=\clr(\bm m)\mid Z=0,\bm X=\bm x]\Pr(\bm X=\bm x)\\
=&\sum_{l,\clr(\bm m), \bm x}\Big[E\left\{Y\{1, \clr(\bm m)\}\mid Z=1,L=l,\bm X=\bm x\right\}\Pr(L=l\mid Z=1,\bm X=\bm x)\\
&\qquad\times \Pr[\clr \{\bm M(1)\}=\clr(\bm m)\mid Z=1,\bm X=\bm x]\Pr(\bm X=\bm x)\Big]\\
&-\sum_{l,\clr(\bm m), \bm x}\Big[E\left\{Y\left(1, \clr(\bm m)\right)\mid Z=1,L=l,\bm X=\bm x\right\}\Pr(L=l\mid Z=1,\bm X=\bm x)\\
&\qquad\times \Pr[\clr\{\bm M(0)\}=\clr(\bm m)\mid Z=0,\bm X=\bm x]\Pr(\bm X=\bm x)\Big]\\
=&\sum_{l,\clr(\bm m), \bm x}\Big[E\left\{Y\left(1, \clr(\bm m)\right)\mid Z=1,L=l,\clr(\bm M)=\clr(\bm m),\bm X=\bm x\right\}\Pr(L=l\mid Z=1,\bm X=\bm x)\\
&\qquad\times \Pr[\clr \{\bm M(1)\}=\clr(\bm m)\mid Z=1,\bm X=\bm x]\Pr(\bm X=\bm x)\Big]\\
&-\sum_{l,\clr(\bm m), \bm x}\Big[E\big\{Y\left(1, \clr(\bm m)\right)\mid Z=1,L=l,\clr(\bm M)=\clr(\bm m),\bm X=\bm x\big\}\Pr(L=l\mid Z=1,\bm X=\bm x)\\
&\qquad\times \Pr[\clr\{\bm M(0)\}=\clr(\bm m)\mid Z=0,\bm X=\bm x]\Pr(\bm X=\bm x)\Big]\\
=&\sum_{l,\clr(\bm m), \bm x}\Big[E\{Y\mid Z=1,L=l,\clr(\bm M)=\clr(\bm m),\bm X=\bm x\}\Pr(L=l\mid Z=1,\bm X=\bm x)\\
&\qquad\times \Pr\{\clr(\bm M)=\clr(\bm m)\mid Z=1,\bm X=\bm x\}\Pr(\bm X=\bm x)\Big]\\
&-\sum_{l,\clr(\bm m), \bm x}\Big[E\{Y\mid Z=1,L=l,\clr(\bm M)=\clr(\bm m),\bm X=\bm x\}\Pr(L=l\mid Z=1,\bm X=\bm x)\\
&\qquad\times \Pr\{\clr(\bm M)=\clr(\bm m)\mid Z=0,\bm X=\bm x\}\Pr(\bm X=\bm x)\Big]\\
=&\sum_{l,\clr(\bm m), \bm x}\Big[E\left\{Y\mid Z=1,L=l,\clr(\bm M)=\clr(\bm m),\bm X=\bm x\right\}\Pr(L=l\mid Z=1,\bm X=\bm x)\\
&\times \left\{\Pr\left(\clr(\bm M)=\clr(\bm m)\mid Z=1,\bm X=\bm x\right)-\Pr\left(\clr(\bm M)=\clr(\bm m)\mid Z=0,\bm X=\bm x\right)\right\} \Pr(\bm X=\bm x)\Big].
\end{flalign*}
\end{small}
\end{proof}

\subsection{ Proof of Theorem 1 in the main manuscript}
\label{s:proofthm1}

\begin{proof}
Based on Proposition \ref{prop:vanderweele}, and under Assumptions  \ref{assump:1}--\ref{assump:3} and \ref{assump:6},
\begin{small}
\begin{equation}
\begin{aligned}
	{\rm IDE} 
	=& \sum\limits_{l,\clr(\bm m), \bm x} \left[E\{Y\mid z_1, l, \clr(\bm m), \bm x\} P(l\mid z_1, \bm x) - E\{Y\mid z_0, l, \clr(\bm m), \bm x\}P(l\mid z_0, \bm x)\right] P\{\clr(\bm m)\mid z_0, \bm x\} P(\bm x);\\ 
   {\rm IIE}
		=& \sum\limits_{l,\clr(\bm m), \bm x} E\{Y\mid z_1, l, \clr(\bm m), \bm x\} P(l\mid z_1, \bm x) \left[P\{\clr(\bm m)\mid z_1,\bm x\} - P\{\clr(\bm m)\mid z_0, \bm x\}\right] P(\bm x).
		\label{eqn:identify22}
		\end{aligned}
\end{equation}
\end{small}
Let 
\begin{flalign*}
    \eta_1 &= \sum\limits_{l,\clr(\bm m), \bm x} E\{Y\mid z_1, l, \clr(\bm m), \bm x\} P(l\mid z_1, \bm x)   P\{\clr(\bm m)\mid z_0, \bm x\} P(\bm x); \\
     \eta_2 &= \sum\limits_{l,\clr(\bm m), \bm x} E\{Y\mid z_0, l, \clr(\bm m), \bm x\} P(l\mid z_0, \bm x)   P\{\clr(\bm m)\mid z_0, \bm x\} P(\bm x); \\
     \eta_3 &= \sum\limits_{l,\clr(\bm m), \bm x} E\{Y\mid z_1, l, \clr(\bm m), \bm x\} P(l\mid z_1, \bm x)   P\{\clr(\bm m)\mid z_1,\bm  x\} P(\bm x). \\
\end{flalign*}

First, we  show that
\begin{equation}
\eta_1 = E\left[\dfrac{\mathbb{I}(Z=z_0) Y}{\Pr(Z=z_0\mid \bm X)} \dfrac{E\{Y\mid Z=z_1, L, \clr(\bm M), \bm X\} \Pr(L\mid Z=z_1, \bm X)}{E\{Y\mid Z=z_0, L, \clr(\bm M), \bm X\} \Pr\{L\mid Z=z_0, \clr(\bm M), \bm X\}}  \right].
\label{eq:eta1}
\end{equation}

\begin{small}
\begin{flalign*}
   & \text{RHS of}\, \eqref{eq:eta1}\\
    =& E_{Z, L,\clr(\bm M), \bm X} E\left[ \left.\dfrac{\mathbb{I}(Z=z_0) Y}{\Pr(Z=z_0\mid \bm X)} \dfrac{E\{Y\mid Z=z_1, L,\clr(\bm M), \bm X\} \Pr(L\mid Z=z_1, \bm X)}{E\{Y\mid Z=z_0, L, \clr(\bm M), \bm X\} \Pr\{L\mid Z=z_0, \clr(\bm M), \bm X\}}    \right|  Z,L,\clr(\bm M), \bm X \right] \\
     =&  \sum\limits_{l,\clr(\bm m), \bm x}  \Bigg[E\left\{  \left.\dfrac{ Y}{\Pr(Z=z_0\mid \bm X=\bm x)} \dfrac{E\{Y\mid z_1, l, \clr(\bm m), \bm x\} P(l\mid z_1, \bm x)}{E\{Y\mid z_0, l, \clr(\bm m), \bm x\} P\{l\mid z_0, \clr(\bm m), \bm x\}}     \right| Z=z_0,l,\clr(\bm m), \bm x \right\} \\
     &\quad\qquad\times P(\bm x) P(z_0\mid \bm x) P\{l,\clr(\bm m)\mid z_0,\bm x\}\Bigg] \\
     =&  \sum\limits_{l,\clr(\bm m), \bm x}  \dfrac{ E\{Y\mid z_0,l,\clr(\bm m),\bm x\}}{\Pr(Z=z_0\mid \bm X=\bm x)} \dfrac{E\{Y\mid z_1, l, \clr(\bm m), \bm x\} P(l\mid z_1, \bm x)}{E\{Y\mid z_0, l, \clr(\bm m), \bm x\} P\{l\mid z_0, \clr(\bm m), \bm x\}}      P(\bm x) P(z_0\mid \bm x) P\{l,\clr(\bm m)\mid z_0,\bm x\} \\
     =& \sum\limits_{l,\clr(\bm m), \bm x}  E\{Y\mid z_1,l,\clr(\bm m), \bm x\}P(l\mid z_1,\bm x) P\{\clr(\bm m)\mid z_0, \bm x\} P(\bm x) \\
     =&  \text{LHS of}\, \eqref{eq:eta1}.
\end{flalign*}
\end{small}

Second, we  show that 
\begin{equation}
    \eta_2 = E\left[\dfrac{\mathbb{I}(Z=z_0) Y}{\Pr(Z=z_0\mid \bm X)} \dfrac{E\{Y\mid Z=z_0, L, \clr(\bm M), \bm X\} \Pr(L\mid Z=z_0, \bm X)}{E\{Y\mid Z= z_0, L, \clr(\bm M), \bm X\} \Pr \{L\mid Z=z_0, \clr(\bm M), \bm X\}}  \right].
    \label{eq:eta2}
\end{equation}

\begin{small}
\begin{flalign*}
     &\text{RHS of}\, \eqref{eq:eta2}\\
     =& E_{Z,L,\clr(\bm M),\bm X} E\left[  \left.\dfrac{\mathbb{I}(Z=z_0) Y}{\Pr(Z=z_0\mid \bm X)} \dfrac{E\{Y\mid Z=z_0, L, \clr(\bm M), \bm X\} \Pr(L\mid Z=z_0, \bm X)}{E\{Y\mid Z=z_0, L, \clr(\bm M), \bm X\} \Pr\{L\mid Z=z_0, \clr(\bm M), \bm X\}}    \right| Z, L,\clr(\bm M), \bm X\right] \\
     =&  \sum\limits_{l,\clr(\bm m), \bm x} \Bigg[ E\left\{  \left.\dfrac{ Y}{\Pr(Z=z_0\mid \bm X=\bm x)} \dfrac{E\{Y\mid z_0, l, \clr(\bm m), \bm x\} P(l\mid z_0, \bm x)}{E\{Y\mid z_0, l, \clr(\bm m), \bm x\} P\{l\mid z_0, \clr(\bm m), \bm x\}}     \right| \bm x,Z=z_0,l,\clr(\bm m) \right\}\\
     &\quad\qquad \times P(\bm x) P(z_0\mid \bm x) P\{l,\clr(\bm m)\mid z_0,\bm x\} \Bigg]\\
     =&  \sum\limits_{l,\clr(\bm m),\bm x}  \dfrac{ E\{Y\mid z_0,l,\clr(\bm m),\bm x\}}{\Pr(Z=z_0\mid \bm X=\bm x)} \dfrac{E\{Y\mid z_0, l, \clr(\bm m), \bm x\} P(l\mid z_0, \bm x)}{E\{Y\mid z_0, l, \clr(\bm m), \bm x\} P(l\mid z_0, \clr(\bm m), \bm x)}      P(\bm x) P(z_0\mid \bm x) P\{l,\clr(\bm m)\mid z_0,\bm x\} \\
     =& \sum\limits_{l,\clr(\bm m),\bm x}  E\{Y\mid z_0,l,\clr(\bm m),\bm x\} P(l\mid z_0,\bm x)P\{\clr(\bm m)\mid z_0, \bm x\} P(\bm x)  \\
     =& \text{LHS of}\, \eqref{eq:eta2}.
\end{flalign*}
\end{small}

Third, we  prove that 
\begin{equation}
    \eta_3 = E\left[\dfrac{\mathbb{I}(Z=z_1) Y}{\Pr(Z=z_1\mid \bm X)} \dfrac{ \Pr(L\mid Z=z_1, \bm X)}{ \Pr\{L\mid Z=z_1, \clr(\bm M), \bm X\}}  \right].
    \label{eq:eta3}
\end{equation}

\begin{small}
\begin{flalign*}
     &\text{RHS of}\, \eqref{eq:eta3}\\
     =& E_{Z,L,\clr(\bm M),\bm X} E\left[ \left.\dfrac{\mathbb{I}(Z=z_1) Y}{\Pr(Z=z_1\mid \bm X)} \dfrac{E\{Y\mid Z=z_1, L, \clr(\bm M), \bm X\} \Pr(L\mid Z=z_1, \bm X)}{E\{Y\mid Z=z_1, L, \clr(\bm M), \bm X\} \Pr\{L\mid Z=z_1, \clr(\bm M), \bm X\}}    \right|Z, L,\clr(\bm M), \bm X \right] \\
     =&  \sum\limits_{l,\clr(\bm m), \bm x}  \Bigg[E\left\{  \left.\dfrac{ Y}{\Pr(Z=z_1\mid \bm X=\bm x)} \dfrac{E\{Y\mid z_1, l, \clr(\bm m), \bm x\} P(l\mid z_1, \bm x)}{E\{Y\mid z_1, l, \clr(\bm m), \bm x\} P\{l\mid z_1, \clr(\bm m), \bm x\}}     \right| \bm x,Z=z_1,l,\clr(\bm m) \right\}\\
     &\quad \qquad \times P(\bm x) P(z_1\mid \bm x) P\{l,\clr(\bm m)\mid z_1,\bm x\}\Bigg] \\
     =&  \sum\limits_{l,\clr(\bm m), \bm x}  \dfrac{ E\{Y\mid z_1,l,\clr(\bm m), \bm x\}}{\Pr(Z=z_1\mid \bm X=\bm x)} \dfrac{E\{Y\mid z_1, l, \clr(\bm m), \bm x\}P(l\mid z_1, \bm x)}{E\{Y\mid z_1, l, \clr(\bm m), \bm x\} P\{l\mid z_1, \clr(\bm m), \bm x\}}      P(\bm x) P(z_1\mid \bm x) P\{l,\clr(\bm m)\mid z_1,\bm x\} \\
     =& \sum\limits_{l,\clr(\bm m), \bm x}  E\{Y\mid z_1,l,\clr(\bm m), \bm x\} P\{\clr(\bm m)\mid z_1, \bm x\} P(\bm x) P(l\mid z_1,\bm x)\\
     =& \text{LHS of}\, \eqref{eq:eta3}.
\end{flalign*}
\end{small}

Therefore, based on \eqref{eqn:identify22},

\begin{flalign*}
{\rm IDE} =&\eta_1-\eta_2\\
=&E\left[\dfrac{\mathbb{I}(Z=z_0) Y}{\Pr(Z=z_0\mid \bm X)} \dfrac{E\{Y\mid Z=z_1, L, \clr(\bm M), \bm X\} \Pr(L\mid Z=z_1, \bm X)}{E\{Y\mid Z=z_0, L, \clr(\bm M), \bm X\} \Pr\{L\mid Z=z_0, \clr(\bm M), \bm X\}}  \right]\\
&- E\left[\dfrac{\mathbb{I}(Z=z_0) Y}{\Pr(Z=z_0\mid \bm X)} \dfrac{ \Pr(L\mid Z=z_0, \bm X)}{ \Pr \{L\mid Z=z_0, \clr(\bm M), \bm X\}}  \right];
\end{flalign*}

\begin{flalign*}
{\rm IIE} =& \eta_3-\eta_1\\
=& E\left[\dfrac{\mathbb{I}(Z=z_1) Y}{\Pr(Z=z_1\mid \bm X)} \dfrac{ \Pr(L\mid Z=z_1, \bm X)}{ \Pr\{L\mid Z=z_1, \clr(\bm M), \bm X\}}  \right]\\
&- E\left[\dfrac{\mathbb{I}(Z=z_0) Y}{\Pr(Z=z_0\mid \bm X)} \dfrac{E\{Y\mid Z=z_1, L, \clr(\bm M), \bm X\}\Pr(L\mid Z=z_1, \bm X)}{E\{Y\mid Z=z_0, L, \clr(\bm M), \bm X\} \Pr\{L\mid Z=z_0, \clr(\bm M), \bm X\}}  \right].
\end{flalign*}
\end{proof}

\subsection{ Proof of Corollary 1 in the main manuscript}
\label{s:proofcoro1}

\begin{proof}

Similar to the proof of Proposition \ref{prop:vanderweele} in Section \ref{s:proofprop2}, and under the assumptions in Corollary 1, we have
\begin{equation*}
\begin{aligned}
		{\rm IDE} 
	=& \sum\limits_{l,\clr(\bm m^{(1)}),\bm x} \Big[\left\{E\left(Y\mid z_1, l, \clr(\bm m^{(1)}), x\right) P(l\mid z_1,\bm x) - E\left(Y\mid z_0, l, \clr(\bm m^{(1)}), \bm x\right)
	P(l\mid z_0, \bm x)\right\}\\
	&\quad \qquad\times P(\clr(\bm m^{(1)})\mid z_0, \bm x) P(\bm x)\Big];\\ 
  {\rm IIE}
		=& \sum\limits_{l,\clr(\bm m^{(1)}),\bm x} E\{Y\mid z_1, l, \clr(\bm m^{(1)}), \bm x\} P(l\mid z_1, \bm x) \left[P\{\clr(\bm m^{(1)})\mid z_1, \bm x\} - P\{\clr(\bm m^{(1)})\mid z_0, \bm x\}\right] P(\bm x).
		\end{aligned}
\end{equation*}

Furthermore, similar to the proof of Theorem 1 in the main manuscript in Section \ref{s:proofthm1}, we have
\begin{flalign*}
{\rm IDE} 
=&E\left[\dfrac{\mathbb{I}(Z=z_0) Y}{\Pr(Z=z_0\mid \bm X)} \dfrac{E\{Y\mid Z=z_1, L, \clr(\bm M^{(1)}), \bm X\} \Pr(L\mid Z=z_1, \bm X)}{E\{Y\mid Z=z_0, L, \clr(\bm M^{(1)}), \bm X\} \Pr\{L\mid Z=z_0, \clr(\bm M^{(1)}), \bm X\}}  \right]\\
&- E\left[\dfrac{\mathbb{I}(Z=z_0) Y}{\Pr(Z=z_0\mid \bm X)} \dfrac{ \Pr(L\mid Z=z_0, \bm X)}{ \Pr \{L\mid Z=z_0, \clr(\bm M^{(1)}), \bm X\}}  \right];
\end{flalign*}

\begin{flalign*}
{\rm IIE} 
=& E\left[\dfrac{\mathbb{I}(Z=z_1) Y}{\Pr(Z=z_1\mid \bm X)} \dfrac{ \Pr(L\mid Z=z_1, \bm X)}{ \Pr\{L\mid Z=z_1, \clr(\bm M^{(1)}), \bm X\}}  \right]\\
&- E\left[\dfrac{\mathbb{I}(Z=z_0) Y}{\Pr(Z=z_0\mid \bm X)} \dfrac{E\{Y\mid Z=z_1, L, \clr(\bm M^{(1)}), \bm X\} \Pr(L\mid Z=z_1, \bm X)}{E\{Y\mid Z=z_0, L, \clr(\bm M^{(1)}), \bm X\} \Pr\{L\mid Z=z_0, \clr(\bm M^{(1)}), \bm X\}}  \right].
\end{flalign*}

 \end{proof}

\end{document}